\documentclass[aps,prd,longbibliography,twocolumn,amsmath,amssymb,showpacs,reprint]{revtex4-1}

\usepackage{graphicx}
\usepackage{dcolumn}
\usepackage{enumitem}
\usepackage{bm}

\usepackage[colorlinks,hyperindex]{hyperref}  
\hypersetup
{
  colorlinks,%
  citecolor=blue,%
  linkcolor=blue,%
  urlcolor=blue,%
}

\begin{document}

\title{
Stueckelberg massive electromagnetism in de Sitter and anti-de Sitter spacetimes: \\
Two-point functions and renormalized stress-energy tensors
}

\author{Andrei Belokogne}
\email{belokogne.andrei@gmail.com}

\affiliation{Equipe Physique
Th\'eorique - Projet COMPA, \\ SPE, UMR 6134 du CNRS
et de l'Universit\'e de Corse,\\
Universit\'e de Corse, BP 52, F-20250 Corte,
France}

\author{Antoine Folacci}
\email{folacci@univ-corse.fr}

\affiliation{Equipe Physique
Th\'eorique - Projet COMPA, \\ SPE, UMR 6134 du CNRS
et de l'Universit\'e de Corse,\\
Universit\'e de Corse, BP 52, F-20250 Corte,
France}

\author{Julien Queva}
\email{queva@univ-corse.fr}

\affiliation{Equipe Physique
Th\'eorique - Projet COMPA, \\ SPE, UMR 6134 du CNRS
et de l'Universit\'e de Corse,\\
Universit\'e de Corse, BP 52, F-20250 Corte,
France}

\date{\today}

\begin{abstract}

By considering Hadamard vacuum states, we first construct the two-point functions associated with Stueckelberg massive electromagnetism in de Sitter and anti-de Sitter spacetimes.
Then, from the general formalism developed in [A. Belokogne and A. Folacci, Phys. Rev. D \textbf{93}, 044063 (2016)], we obtain an exact analytical expression for the vacuum expectation
value of the renormalized stress-energy tensor of the massive vector field propagating in these maximally symmetric spacetimes.

\end{abstract}


\maketitle

\tableofcontents

\section{Introduction}
\label{Sec.I}

In a recent article, we discussed the covariant quantization of Stueckelberg massive electromagnetism on an arbitrary four-dimensional curved spacetime $(\mathcal{M}, g_{\mu \nu})$ with no boundary and we constructed, for Hadamard quantum states, the expectation value of the renormalized stress-energy tensor (RSET) \cite{Belokogne:2015etf}. Here, we do not return to the motivations leading us to consider Stueckelberg massive electromagnetism in curved spacetime.
The interested reader is invited to consult our previous article, in particular its introduction as well as references therein.
The formalism developed in Ref.~\cite{Belokogne:2015etf} permitted us to discuss, as an application, the Casimir effect outside a perfectly conducting medium with a plane boundary.

In the present paper, we shall address a much more difficult problem which could have interesting implications in cosmology of the very early Universe or in the context of the AdS/CFT correspondence: we shall obtain an exact analytical expression for the vacuum expectation value of the RSET of the massive vector field propagating in de Sitter and anti-de Sitter spacetimes. It is interesting to note that such results do not exist in the literature while the RSETs associated with the massive scalar field and the massive spinor field have been obtained quite a long time ago (see, e.g., Refs.~\cite{Dowker:1975tf,Bunch:1978yq,Bernard:1986vc,Tadaki:1988cn,Camporesi:1991nw,Caldarelli:1998wk,Kent:2014nya} for the case of the massive scalar field and Refs.~\cite{Camporesi:1992wn,Landete:2013axa,Ambrus:2015mfa} for the case of the massive spinor field). In fact, this void in the literature can easily be explained. Indeed, even if there exist numerous works concerning the massive vector field in de Sitter and anti-de Sitter spacetimes \cite{Boerner:1969ff,SchomblondSpindel76,Drechsler:1978cw,Allen:1985wd,Janssen:1986fz,Gazeau:1987nu,Gazeau:1999xn,Tsamis:2006gj,Miao:2010vs,Frob:2013qsa,Narain:2014oja}, the two-point functions are in general constructed in the framework of the de Broglie-Proca theory and, as a consequence, do not display the usual Hadamard singularity (see the last remark in the conclusion of Ref.~\cite{Belokogne:2015etf}) which is a fundamental ingredient of regularization and renormalization techniques in curved spacetime.

In our article, we shall focus on Stueckelberg electromagnetism defined, at the quantum level, by the action \cite{Ruegg:2003ps,Belokogne:2015etf}
\begin{eqnarray}
\label{Action_Stueck_Quant}
&& S\left[A_\mu, \Phi, C, C^\ast, g_{\mu \nu} \right] =
  S_A \left[A_\mu, g_{\mu \nu} \right]
\nonumber \\ && \qquad
+ S_\Phi \left[\Phi, g_{\mu \nu} \right]
+ S_\mathrm{Gh} \left[C, C^\ast, g_{\mu \nu} \right],
\end{eqnarray}
where
\begin{eqnarray}
\label{Action_Stueck_Amu_xi=1}
&& S_A = \int_{\cal M} d^4 x \, \sqrt{-g}  \, \left[
- \frac{1}{4} F^{\mu \nu} F_{\mu \nu}
\right. \nonumber \\ && \left.  \qquad \qquad
- \frac{1}{2} \, m^2 A^\mu A_\mu
- \frac{1}{2} \, \left( \nabla^\mu A_\mu \right)^2
\right]
\end{eqnarray}
denotes the action associated with the massive vector field $A_\mu$ with mass $m$ (here, $F_{\mu \nu} = \nabla_\mu A_\nu - \nabla_\nu A_\mu = \partial_\mu A_\nu - \partial_\nu A_\mu$ is the associated field strength) and
\begin{eqnarray}
\label{Action_Stueck_Quant_Phi}
&& \hspace{-5mm} S_\Phi = \int_{\cal M} d^4 x \, \sqrt{-g} \, \left(
- \frac{1}{2} \, \nabla^\mu \Phi \nabla_\mu \Phi
- \frac{1}{2} \,  m^2 \Phi^2
\right)
\end{eqnarray}
is the action governing the auxiliary Stueckelberg scalar field $\Phi$. The last action term in Eq.~\eqref{Action_Stueck_Quant} is the compensating ghost contribution given by
\begin{equation}
\label{Action_Stueck_Quant_Ghosts}
S_\mathrm{Gh} = \int_{\cal M} d^4 x \, \sqrt{-g} \, \left( \nabla^\mu C^\ast \nabla_\mu C +  m^2 C^\ast C \right),
\end{equation}
where $C$ and $C^\ast$ are two fermionic ghost fields. Here, it is important to note that some authors dealing with Stueckelberg electromagnetism (see, e.g., Refs.~\cite{ItzyksonZuber,Janssen:1986fz,Frob:2013qsa}) have considered Stueckelberg electromagnetism defined from the sole action
\begin{eqnarray}
\label{Action_Stueck_Amu_xi}
&& S_A = \int_{\cal M} d^4 x \, \sqrt{-g} \, \left[
- \frac{1}{4} F^{\mu \nu} F_{\mu \nu}
\right. \nonumber \\ && \left. \qquad \qquad
- \frac{1}{2} \, m^2 A^\mu A_\mu
- \frac{1}{2\xi} \, \left( \nabla^\mu A_\mu \right)^2
\right],
\end{eqnarray}
where $\xi$ is a gauge parameter. In fact, these authors were mainly interested by the determination of the Feynman propagator associated with the massive vector field $A_\mu$. Of course, in order to calculate physical quantities such as the RSET associated with Stueckelberg electromagnetism, the full action must be considered, i.e., it is necessary to take also into account, in addition to the contribution of the massive vector field, those of the auxiliary Stueckelberg field and of the ghost fields. In our article, we shall not consider the case of an arbitrary gauge parameter $\xi$. Indeed, as we have already noted in Ref.~\cite{Belokogne:2015etf}, if we want to work with Hadamard quantum states, it is necessary to take $\xi = 1$. However, it is interesting to recall that Fr{\"o}b and Higuchi in a recent article \cite{Frob:2013qsa} have provided, for an arbitrary value of $\xi$, a mode-sum construction of the two-point functions for the massive vector field by working in the Poincar\'e patch of de Sitter space. Their results have permitted them to recover, as particular cases, the two-point functions obtained by Allen and Jacobson in Ref.~\cite{Allen:1985wd} (they correspond to $\xi \to \infty$, i.e., to the de Broglie-Proca theory) and those obtained by Tsamis and Woodard in Ref.~\cite{Tsamis:2006gj} (they correspond to $\xi \to 0$).

Our article is organized as follows. In Sec.~\ref{Sec.II}, we construct the Wightman functions associated with the massive vector field $A_\mu$, the Stueckelberg auxiliary scalar field $\Phi$ and the ghost fields $C$ and $C^\ast$ and, from them, we deduce by analytic continuation all the other two-point functions. We do not use a mode-sum construction as in Ref.~\cite{Frob:2013qsa}, but we extend the approach of Allen and Jacobson in Ref.~\cite{Allen:1985wd} (see also Refs.~\cite{Dullemond:1984bc,Janssen:1986fz,Folacci:1990ea}). More precisely, by assuming that the vacuum is a maximally symmetric quantum state, we solve the wave equations for the various Wightman functions involved by taking into account, as constraints, two Ward identities; we then fix the remaining integration constants by imposing (i) Hadamard-type singularities at short distance and (ii) in de Sitter spacetime, the regularity of the solutions at the antipodal point or (iii) in anti-de Sitter spacetime, that the solutions fall off as fast as possible at spatial infinity. In Sec.~\ref{Sec.III}, from the general formalism developed in Ref.~\cite{Belokogne:2015etf}, we obtain an exact analytical expression for the vacuum expectation
value of the RSET of the massive vector field propagating in de Sitter and anti-de Sitter spacetimes. The geometrical ambiguities are fixed by considering the flat-space limit and, moreover, we consider the two alternative but equivalent expressions for the renormalized expectation value given in Ref.~\cite{Belokogne:2015etf} in order to discuss the zero-mass limit of our results. Finally, in a conclusion (Sec.~\ref{Sec.IV}), we briefly consider the possible extension of our work to Stueckelberg electromagnetism in an arbitrary $\xi$ gauge.

It should be noted that, in this article, we use units such that $\hbar=c=G=1$ and the geometrical conventions of Hawking and Ellis \cite{HawkingEllis} concerning the definitions of the scalar curvature $R$, the Ricci tensor $R_{\mu \nu}$ and the Riemann tensor $R_{\mu \nu \rho \sigma}$ as well as the commutation of covariant derivatives. Moreover, we will frequently refer to our previous article \cite{Belokogne:2015etf} and we assume that the reader has ``in hand" a copy of it.

\section{Two-point functions of Stueckelberg electromagnetism}
\label{Sec.II}

In this section, we shall construct the various two-point functions involved in Stueckelberg massive electromagnetism from the Wightman functions associated with the vector field $A_\mu$, the Stueckelberg scalar field $\Phi$ and the ghost fields $C$ and $C^\ast$.

\subsection{de Sitter and anti-de Sitter spacetimes}
\label{Sec.IIa}

Here, we gather some results concerning (i) the geometry of the four-dimensional de Sitter spacetime ($\mathrm {dS}^4$) and the four-dimensional anti-de Sitter spacetime ($\mathrm {AdS}^4$) as well as (ii) the properties of some geometrical objects defined on these maximally symmetric gravitational backgrounds. We have minimized the information on these topics (for more details and proofs, see Refs.~\cite{HawkingEllis,Avis:1977yn,Allen:1985ux,Allen:1985wd,Folacci:1990ea}). Those results are necessary to construct the Wightman functions of Stueckelberg electromagnetism and, in Sec.~\ref{Sec.III}, will permit us to simplify in $\mathrm {dS}^4$ and $\mathrm {AdS}^4$ the formalism developed in Ref.~\cite{Belokogne:2015etf}.

$\mathrm {dS}^4$ and $\mathrm {AdS}^4$ are maximally symmetric spacetimes of constant scalar curvature (positive for the former and negative for the latter) which are locally characterized by the relations
\begin{subequations}
\allowdisplaybreaks
\label{Expr_RiemannTensors}
\begin{align}
\label{Expr_RiemannTensor}
& R_{\mu \nu \rho \tau} = (R/12) \, \left( g_{\mu \rho} g_{\nu \tau} - g_{\mu \tau} g_{\nu \rho} \right) ,
\\
\label{Expr_RicciTensor}
& R_{\mu \nu} = (R/4) \, g_{\mu \nu} ,
\end{align}
\noindent and
\begin{align}
\label{Expr_RicciScalar}
& R =
  \begin{cases}
    + 12 H^2   & \quad \text{for dS}^4, \\
    - 12 K^2   & \quad \text{for AdS}^4.\\
  \end{cases}
\end{align}
\end{subequations}
Here $H$ and $K$ are two positive constants of dimension $(\mathrm{length})^{-1}$. The relations (\ref{Expr_RiemannTensor})--(\ref{Expr_RicciScalar}) are useful in order to simplify the various covariant Taylor series expansions involved in the Hadamard renormalization process (see Ref.~\cite{Belokogne:2015etf}) and will be extensively used in Sec.~\ref{Sec.III}.

$\mathrm {dS}^4$ and $\mathrm {AdS}^4$ can be realized as the four-dimensional hyperboloids
\begin{equation}
\label{Hyperboloid_in_MinkowskiSpace}
\eta_{a b} X^a X^b = 12 / R
\end{equation}
embedded in the flat five-dimensional space $\mathbb{R}^5$ equipped with the metric
\begin{equation}
\label{MinkowskiMetric}
\eta_{a b} =
  \begin{cases}
    \mathrm{diag}(-1, +1, +1, +1, +1)   & \quad \text{for dS}^4,\\
    \mathrm{diag}(-1, -1, +1, +1, +1)   & \quad \text{for AdS}^4.\\
  \end{cases}
\end{equation}
Equations~\eqref{Hyperboloid_in_MinkowskiSpace} and \eqref{MinkowskiMetric} make it obvious that $O(1,4)$ is the symmetry group of $\mathrm {dS}^4$ and that its topology is that of $\mathbb{R} \times S^3$, while $O(2,3)$ is the symmetry group of $\mathrm {AdS}^4$ whose topology is that of $S^1 \times \mathbb{R}^3$. It is important to recall that, in order to avoid closed timelike curves in $\mathrm {AdS}^4$, it is necessary to ``unwrap'' the circle $S^1$ to go onto its universal covering space $\mathbb{R}^1$ and then $\mathrm {AdS}^4$ has the topology of $\mathbb{R}^4$.

In the context of field theories in curved spacetime, the geodetic interval $\sigma(x,x')$, defined as one-half the square of the geodesic distance between the points $x$ and $x'$, is of fundamental interest (see, e.g., Refs.~\cite{DeWitt:1960fc,Birrell:1982ix}) and the Hadamard renormalization process developed in Ref.~\cite{Belokogne:2015etf}, which we will exploit in Sec.~\ref{Sec.III}, is based on an extensive use of this geometrical object. However, in this section, even though $\sigma(x,x')$ is invariant under the symmetry group of $\mathrm {dS}^4$ or $\mathrm {AdS}^4$, it is advantageous to consider instead the real quadratic form
\begin{equation}
\label{Def_z_XX}
z(x,x') = \frac{1}{2} \, \left[ 1 + (R/12) \, \eta_{a b} X^a(x) X^b(x') \right]
\end{equation}
in order to construct the two-point functions of Stueckelberg electromagnetism. In Eq.~\eqref{Def_z_XX}, $X^a(x)$ and $X^b(x')$ are the coordinates of the points $x$ and $x'$ on the hyperboloid (\ref{Hyperboloid_in_MinkowskiSpace}) defining $\mathrm {dS}^4$ or $\mathrm {AdS}^4$ and $\eta_{a b}$ is the corresponding metric given by (\ref{MinkowskiMetric}). This quadratic form is obviously invariant under the symmetry group of $\mathrm {dS}^4$ or $\mathrm {AdS}^4$ and is moreover defined on the whole spacetime, while $\sigma(x,x')$ is not defined everywhere because there is not always a geodesic between two arbitrary points in these maximally symmetric spacetimes. However, when $\sigma(x,x')$ is defined, we have
\begin{subequations}
\allowdisplaybreaks
\label{Def_z}
\begin{align}
\label{Def_z_Ra}
z(x,x') &= \frac{1}{2} \, \left[ 1 + \cos \sqrt{ (R/6) \, \sigma(x,x') } \right]
 \\ \label{Def_z_Rb}
&= \cos^2 \sqrt{ (R/24) \, \sigma(x,x') }
\end{align}
\end{subequations}
or, equivalently,
\begin{equation}
\label{Def_z_H}
z(x,x') = \cos^2 \sqrt{ (H^2/2) \, \sigma(x,x') }
\end{equation}
in $\mathrm {dS}^4$ and
\begin{equation}
\label{Def_z_K}
z(x,x') = \cosh^2 \sqrt{ (K^2/2) \, \sigma(x,x') }
\end{equation}
in $\mathrm {AdS}^4$. The previous relations can be inverted and used to define $\sigma(x,x')$ globally. In fact, it will chiefly help us, in Sec.~\ref{Sec.III}, to reexpress the two-point functions obtained here in terms of $\sigma(x,x')$.

\begin{figure}
\includegraphics[width=80mm, trim={0mm 21mm 0mm 21mm}]{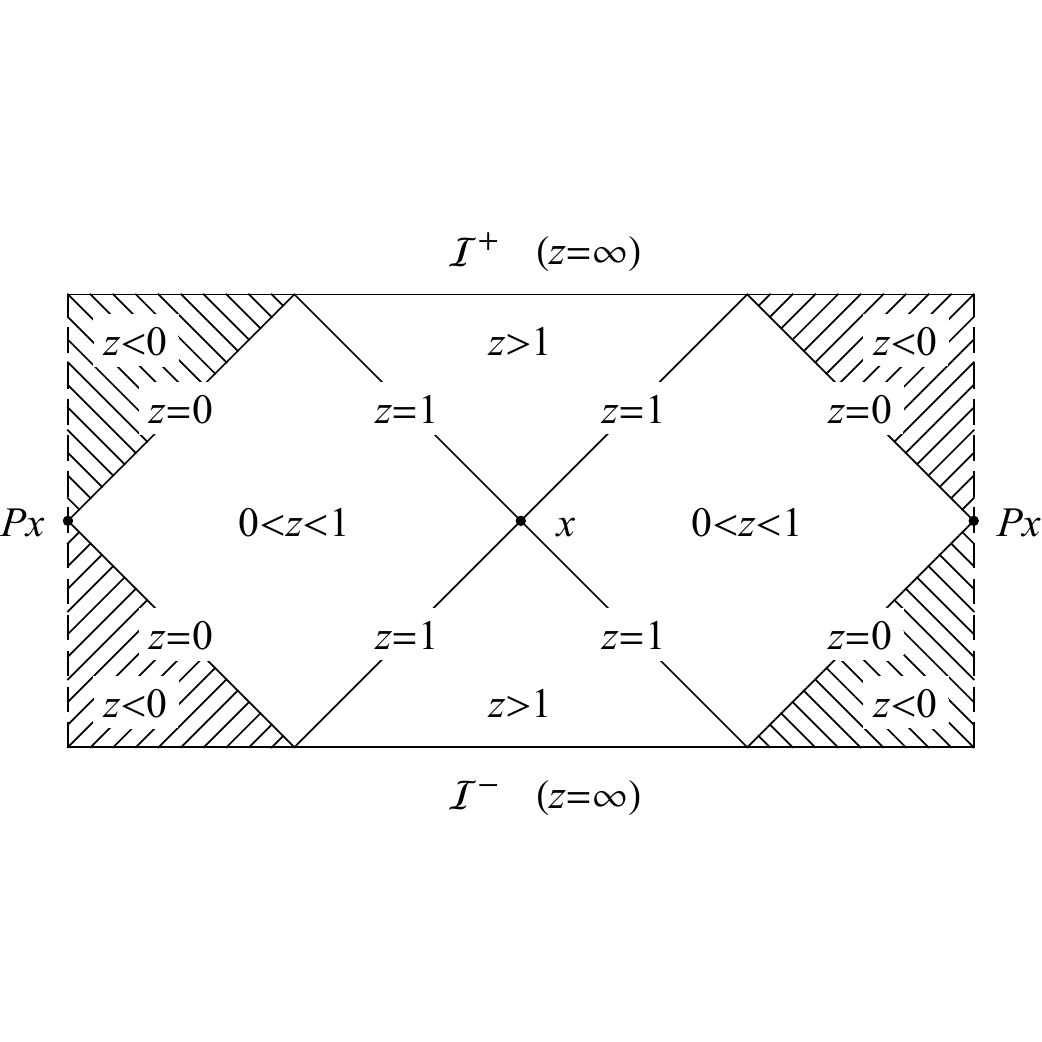} 
    \caption{Carter-Penrose diagram of $\mathrm {dS}^4$. Without loss of generality, the point $x$ can be taken to be any point in spacetime. $Px$ denotes its associated antipodal point. The left and right edges of this diagram must be identified along the dashed lines. $\mathcal{I}^+$ and $\mathcal{I}^-$ denote, respectively, the future and past spacelike infinities for timelike and null geodesics. The hatched area is the set of points $x'$ which cannot be reached by geodesics from $x$ and for which $\sigma (x,x')$ is not defined.}
    \label{fig_dS_PenroseDiagram}
\end{figure}

\begin{figure}
\includegraphics[width=80mm]{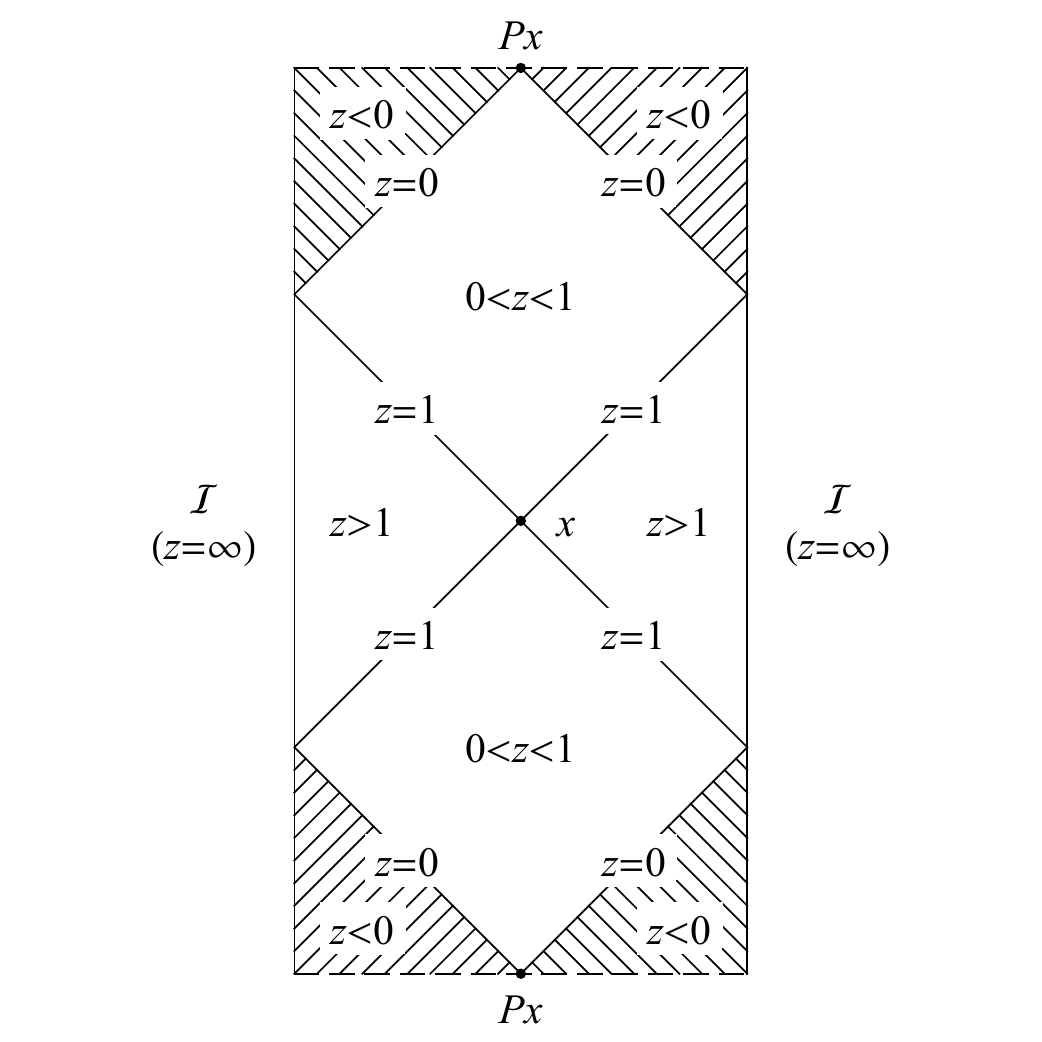}
    \caption{Carter-Penrose diagram of $\mathrm {AdS}^4$. Without loss of generality, the point $x$ can be taken to be any point in spacetime. $Px$ denotes its associated antipodal point. Here, we only display an elementary cell of the universal covering of anti-de Sitter spacetime. The whole manifold is obtained by gluing along the dashed lines the top edge of one cell with the bottom edge of another one and by replicating, ad nauseam, this process. $\mathcal{I}$ denotes timelike infinity. The hatched area is the set of points $x'$ which cannot be reached by geodesics from $x$ and for which $\sigma (x,x')$ is not defined.}
    \label{fig_AdS_PenroseDiagram}
\end{figure}

We shall now point out some useful properties of $z(x,x')$. With respect to the antipodal transformation which sends the point $x$ with coordinates $X^a(x)$ on the hyperboloid (\ref{Hyperboloid_in_MinkowskiSpace}) to its antipodal point $Px$ with coordinates
\begin{equation}
\label{AntipodalTransformation_X}
X^a(Px) = - X^a(x),
\end{equation}
we have
\begin{equation}
\label{AntipodalTransformation_z}
z(x,Px') = 1 - z(x,x').
\end{equation}
We have also
\begin{subequations}
\label{Range_z_dS}
\begin{align}
& z(x,x')>1    \,\, \text{if $x$ and $x'$ are timelike related} ,
\\
& z(x,x')=1    \,\, \text{if $x$ and $x'$ are null related} ,
\\
& 0< z(x,x')<1    \,\, \text{if $x$ and $x'$ are spacelike related},
\\
& z(x,x') \leq 0  \,\, \text{if $x$ and $x'$ cannot be joined by a geodesic},
\end{align}
\end{subequations}
in $\mathrm {dS}^4$ and
\begin{subequations}
\label{Range_z_AdS}
\begin{align}
& z(x,x')>1     \,\,  \text{if $x$ and $x'$ are spacelike related},
\\
& z(x,x')=1     \,\, \text{if $x$ and $x'$ are null related},
\\
& 0<z(x,x')<1   \,\, \text{if $x$ and $x'$ are timelike related},
\\
& z(x,x') \leq 0  \,\, \text{if $x$ and $x'$ cannot be joined by a geodesic},
\end{align}
\end{subequations}
in $\mathrm {AdS}^4$. All these results can be visualized in the Carter-Penrose diagrams of $\mathrm {dS}^4$ (see Fig.~\ref{fig_dS_PenroseDiagram}) and $\mathrm {AdS}^4$ (see Fig.~\ref{fig_AdS_PenroseDiagram}).

To conclude this subsection, we recall that in $\mathrm {dS}^4$ and $\mathrm {AdS}^4$, any bitensor which is invariant under the spacetime symmetry group (a maximally symmetric tensor) can be expressed only in terms of the bitensors $z$, $z_{;\mu}$, $z_{;\mu'}$, $g_{\mu \nu}$, $g_{\mu' \nu'}$ and  $g_{\mu \nu'}$ \cite{Allen:1985wd}. Here, $g_{\mu \nu'}$ is the usual bivector of parallel transport from $x$ to $x'$ which is defined by the differential equation $g_{\mu \nu' ;\rho} \sigma^{;\rho} = 0$ and the boundary condition $\lim_{x' \to x} g_{\mu \nu'} (x,x') = g_{\mu \nu} (x)$. For example, a two-point function $G(x,x')$ associated with a scalar field is necessarily of the form $G(z)$ and a two-point function $G_{\mu \nu'}(x,x')$ associated with a vector field can be written in the form $A(z)g_{\mu \nu'}+B(z)z_{;\mu}z_{;\nu'}$. This very important remark will simplify the construction of the two-point functions of Stueckelberg electromagnetism.
Besides, in order to handle these functions in connection with the wave equations and the Ward identities, it will be necessary to use the following geometrical relations
\begin{subequations}
\label{GeometricalRelations}
\begin{align}
\allowdisplaybreaks
& z_{;\mu} z^{;\mu} = (R/12) \, z (1-z) , \label{GeometricalRelations_a}
\\
& g_{\mu \nu' ;\rho} z^{;\rho} = 0 ,
\\
& g_{\mu \nu'} z^{;\mu} = - z_{;\nu'} ,
\\
& g_{\mu \nu'} z^{;\nu'} = - z_{;\mu} ,
\\
& z_{;\mu \nu} = (R/24) \, (1-2z) g_{\mu \nu} ,
\\
& z_{;\mu \nu'} = (R/24) \, g_{\mu \nu'} + (1/z) \, z_{;\mu} z_{;\nu'} ,
\\
& g_{\mu \nu' ;\rho} = - (1/z) \left( g_{\mu \rho} z_{;\nu'} + g_{\rho \nu'} z_{;\mu} \right) ,
\\
& g_{\mu \nu' ;\rho'} = - (1/z) \left( g_{\mu \rho'} z_{;\nu'} + g_{\nu' \rho'} z_{;\mu} \right) ,
\\
& g_{\mu \nu' ;\rho}^{\phantom{\mu \nu' ;\rho};\rho} = (R/12) \, [(z-1)/z] \, g_{\mu \nu'} + (2/z^2) \, z_{;\mu} z_{;\nu'},
\end{align}
\end{subequations}
and to note that the d'Alembertian operator acting on biscalar functions is given by
\begin{equation}
\label{Operator_dAlembertian}
\Box = \frac{R}{12} \left[ z (1-z) \, \frac{d^2}{dz^2} + (2-4z) \, \frac{d}{dz} \right] .
\end{equation}

\subsection{Stueckelberg theory, wave equations and Ward identities for the Wightman functions}
\label{Sec.IIb}

From the quantum action (\ref{Action_Stueck_Quant}), we can easily obtain the wave equations satisfied by the massive vector field $A_\mu$, the auxiliary scalar field $\Phi$ and the ghost fields $C$ and $C^\ast$. They have been derived in our previous article [see Eqs.~(19)--(21) in Ref.~\cite{Belokogne:2015etf}] and, in $\mathrm {dS}^4$ or in $\mathrm {AdS}^4$, due to Eq.~\eqref{Expr_RicciTensor}, they reduce to
\begin{subequations}
\label{WaveEq_A_Phi_Gh}
\begin{align}
\label{WaveEq_A}
& \left( \Box - R/4 - m^2 \right) A_\mu = 0 ,
\\
\label{WaveEq_Phi}
& \left( \Box - m^2 \right) \Phi = 0 ,
\\
\label{WaveEq_Gh1}
& \left( \Box - m^2 \right) C = 0 ,
\\
\label{WaveEq_Gh2}
&\left( \Box - m^2 \right) C^\ast = 0 .
\end{align}
\end{subequations}

From now on, we shall assume that the Stueckelberg theory is quantized in a normalized vacuum state $|0\rangle$ and, in addition, that this quantum state is (i) maximally symmetric and (ii) of Hadamard type. We recall that, in the context of the calculation of the renormalized expectation value of the stress-energy-tensor operator with respect to a vacuum $|0\rangle$, it is convenient to work with the Feynman propagators or, equivalently, with the so-called Hadamard Green functions associated with the fields of the theory \cite{Belokogne:2015etf}. So, we need in $\mathrm {dS}^4$ and $\mathrm {AdS}^4$ the explicit expressions of these two-point functions. In fact, it is possible to construct the zoo of the two-point functions of the theory from the Wightman functions and, in a first step, we shall focus on these particular two-point functions.

We recall that the Wightman function associated with the massive vector field $A_\mu$ is given by
\begin{align}
\label{Wightman_A}
G^{(+)A}_{\mu \nu'} (x,x') = \langle 0 |  A_\mu (x)  A_{\nu'} (x') | 0 \rangle
\end{align}
and satisfies the wave equation [see also Eq.~\eqref{WaveEq_A}]
\begin{align}
\label{WaveEq_WA}
\left( \Box_x - R/4  - m^2 \right) G^{(+)A}_{\mu \nu'} (x,x') = 0 .
\end{align}
Similarly, the Wightman function associated with the auxiliary scalar field $\Phi$ is given by
\begin{align}
\label{Wightman_Phi}
G^{(+)\Phi} (x,x') = \langle 0 | \Phi (x)  \Phi (x') | 0 \rangle
\end{align}
and is a solution of [see also Eq.~\eqref{WaveEq_Phi}]
\begin{align}
\label{WaveEq_WPhi}
\left( \Box_x  - m^2 \right) G^{(+)\Phi} (x,x') = 0 ,
\end{align}
while the Wightman function associated with the ghost fields is defined by
\begin{align}
\label{Wightman_Gh}
G^{(+)\mathrm{Gh}} (x,x') = \langle 0 |  C^\ast (x) C (x')  | 0 \rangle
\end{align}
and satisfies the wave equation [see also Eq.~\eqref{WaveEq_Gh2}]
\begin{align}
\label{WaveEq_WGh}
\left( \Box_x  - m^2 \right) G^{(+)\mathrm{Gh}} (x,x') = 0 .
\end{align}
Moreover, we have two Ward identities that relate these three Wightman functions. We can write
\begin{align}
\label{WardId_A_Gh}
\nabla^\mu G^{(+)A}_{\mu \nu'} (x,x') + \nabla_{\nu'} G^\mathrm{(+)\mathrm{Gh}} (x,x') = 0
\end{align}
and
\begin{align}
\label{WardId_Phi_Gh}
G^{(+)\Phi} (x,x') - G^{(+)\mathrm{Gh}} (x,x') = 0 .
\end{align}
It should be noted that, in our previous article, we have derived the Ward identities for the Feynman propagators [see Eqs.~(33) and (29) in Ref.~\cite{Belokogne:2015etf}] and for the Hadamard Green functions [see Eqs.~(65) and (66) in Ref.~\cite{Belokogne:2015etf}]. Here we have written them for the Wightman functions. They can be derived in the same way, i.e., from the wave equations by using arguments of uniqueness \cite{DeWitt:1960fc}. The Ward identity (\ref{WardId_Phi_Gh}) expresses the equality of the Wightman functions associated with the auxiliary scalar field and the ghost fields. So, thereafter we shall omit the labels $\Phi$ and $\mathrm{Gh}$ and use a generic form for these two Green functions by writing
\begin{align}
\label{WGreenFn}
G^{(+)} (x,x') = G^{(+)\Phi} (x,x') = G^{(+)\mathrm{Gh}} (x,x') .
\end{align}

\subsection{Explicit expression for the Wightman functions in $\mathrm {dS}^4$ and $\mathrm {AdS}^4$}
\label{Sec.IIc}

\subsubsection{General form of the Wightman functions in maximally symmetric backgrounds}
\label{Sec.IIc1}

We have previously assumed that the vacuum $|0\rangle$ is a maximally symmetric state. As a consequence, we can express the Wightman functions $G^{(+)A}_{\mu \nu'} (x,x')$ and $G^{(+)} (x,x') $ as a function of the quadratic form $z(x,x')$ [see also the last paragraph of the subsection \ref{Sec.IIa}] and write
\begin{align}
\label{ScalarWFn_z}
G^{(+)}(x,x') = G^{(+)}(z)
\end{align}
for the scalar Wightman functions (\ref{Wightman_Phi}) and (\ref{Wightman_Gh}) and
\begin{align}
\label{VectorWFn_z}
G^{(+)A}_{\mu \nu'}(x,x') = G^{(+)A}_{\mu \nu'}(z) = \alpha(z) \, g_{\mu \nu'} + 4 \, \beta(z) \, z_{;\mu} z_{;\nu'}
\end{align}
for the vector Wightman function (\ref{Wightman_A}).

By inserting (\ref{ScalarWFn_z}) into the wave equation (\ref{WaveEq_WPhi}) or (\ref{WaveEq_WGh}) and by taking into account the relations (\ref{GeometricalRelations}) and (\ref{Operator_dAlembertian}), we obtain the differential equation
\begin{align}
\label{WaveEq_G_z}
\left[ z(1-z) \, \frac{d^2}{dz^2} + (2-4z) \, \frac{d}{dz} - \frac{12m^2}{R} \right] G^{(+)}(z) = 0.
\end{align}
Similarly, inserting (\ref{VectorWFn_z}) into the wave equation (\ref{WaveEq_WA}) leads to a system of two coupled differential equations for the functions $\alpha(z)$ and $\beta(z)$ given by
\begin{subequations}
\allowdisplaybreaks
\label{WaveEq_alpha_beta_z}
\begin{align}
\label{WaveEq_alpha_z}
&\left[ z(1-z) \, \frac{d^2}{dz^2} + (2-4z) \, \frac{d}{dz} - \frac{2z+1}{z} - \frac{12m^2}{R} \right] \alpha(z)
\nonumber \\
&\qquad + \frac{R}{6} \, (1-2z) \, \beta(z) = 0
\end{align}
and
\begin{align}
\label{WaveEq_beta_z}
&\left[ z(1-z) \, \frac{d^2}{dz^2} + (4-8z) \, \frac{d}{dz} - \frac{10z-1}{z} - \frac{12m^2}{R} \right] \beta(z)
\nonumber \\
&\qquad + \frac{6}{R} \, \frac{1}{z^2} \, \alpha(z) = 0,
\end{align}
\end{subequations}
while, from the Ward identity \eqref{WardId_A_Gh}, we obtain
\begin{align}
\label{WardId_alpha_beta_z}
&\left[ \frac{d}{dz} + \frac{3}{z} \right] \alpha(z) - \left[ \frac{R}{3} \, z(1-z) \, \frac{d}{dz} + \frac{5R}{6} \, (1-2z) \right] \beta(z)
\nonumber \\
&\qquad - \frac{d}{dz} G^{(+)}(z) = 0 .
\end{align}

Equation \eqref{WaveEq_G_z} is a hypergeometric differential equation of the form \cite{AS65,Olver:2010:NHMF,NIST:DLMF}
\begin{align}
\label{Def_HypegeometricDiffEq}
& \left[ z(1-z) \, \frac{d^2}{dz^2} + [c-(a+b+1)z] \, \frac{d}{dz} \right. \nonumber \\
& \qquad\qquad\qquad \left. \phantom{\frac{d^2}{dz^2}} - a b \right] F(a,b;c;z) = 0
\end{align}
with $a=3/2+\kappa$, $b=3/2-\kappa$ and $c=2$ where
\begin{equation}
\label{Expression_kappa}
\kappa = \sqrt{9/4 - 12 \, m^2/R}.
\end{equation}
This equation is invariant under the transformation $z \to 1-z$ because the parameters $a$, $b$ and $c$ satisfy $a+b+1=2c$. As a consequence, we can write
\begin{align}
\label{Sol_WScalarFn}
& G^{(+)}(x,x') = C^1_G \, F(3/2+\kappa, 3/2-\kappa; 2; z)
\nonumber \\
&\qquad + C^2_G \, F(3/2+\kappa, 3/2-\kappa; 2; 1-z)
\end{align}
where $C^1_G$ and $C^2_G$ are two integration constants.

The differential equations \eqref{WaveEq_alpha_z} and \eqref{WaveEq_beta_z} which provide the Wightman function \eqref{VectorWFn_z} are much more complicated to solve.
In order to do so, we introduce the new function $\gamma(z)$ defined by
\begin{align}
\label{Def_gamma_z}
\gamma(z) = \alpha(z) - (R/3) \, z(1-z) \, \beta(z) ,
\end{align}
and rewrite the Ward identity \eqref{WardId_alpha_beta_z} in the form
\begin{equation}
\label{WardId_gamma_beta_z}
\left[ \frac{d}{dz} + \frac{3}{z} \right] \gamma(z) + \frac{R}{2} \, \beta(z) - \frac{d}{dz} G^{(+)}(z) = 0.
\end{equation}
Then, by using Eqs.~\eqref{Def_gamma_z} and \eqref{WardId_gamma_beta_z}, we can combine the differential equations \eqref{WaveEq_alpha_z} and \eqref{WaveEq_beta_z} and we have
\begin{align}
\label{WaveEq_gamma_z}
&\left[ z(1-z) \, \frac{d^2}{dz^2} + (3-6z) \, \frac{d}{dz} - 6 - \frac{12m^2}{R} \right] \gamma(z) =
\nonumber \\
&\qquad (1-2z) \, \frac{d}{dz} G^{(+)}(z) .
\end{align}
The general solution $\gamma(z)$ of this nonhomogeneous hypergeometric differential equation is the sum
\begin{align}
\label{SolG_gamma_z}
\gamma(z) = \gamma_c(z) + \gamma_p(z)
\end{align}
of the complementary solution $\gamma_c(z)$ and a particular solution $\gamma_p(z)$. $\gamma_c(z)$ is the general solution of the hypergeometric differential equation (\ref{Def_HypegeometricDiffEq}) with $a=5/2+\lambda$, $b=5/2-\lambda$ and $c=3$ where
\begin{equation}
\label{Expression_lambda}
\lambda = \sqrt{1/4 - 12 \, m^2/R}.
\end{equation}
Since the coefficients $a$, $b$ and $c$ fulfill again $a+b+1=2c$, we can therefore write
\begin{align}
\label{SolC_gamma_z}
& \gamma_c(z) = C^1_\gamma \, F(5/2+\lambda, 5/2-\lambda; 3; z)
\nonumber \\ &\qquad
+ C^2_\gamma \, F(5/2+\lambda, 5/2-\lambda; 3; 1-z)
\end{align}
where $C^1_\gamma$ and $C^2_\gamma$ are two new integration constants. Furthermore, it is rather easy to check that it is possible to take as a particular solution of (\ref{WaveEq_gamma_z})
\begin{align}
\label{SolP_gamma_z}
& \gamma_p(z) = \frac{1}{9/4 - \kappa^2}
\nonumber \\ &\qquad
\times \left[ z(1-z) \, \frac{d^2}{dz^2} + (1/2) \, (1-2z) \, \frac{d}{dz} \right] G^{(+)}(z) .
\end{align}
We have now at our disposal the general solution of the nonhomogeneous differential equation (\ref{WaveEq_gamma_z}). This permits us to establish the expression of the Wightman function \eqref{VectorWFn_z} by determining $\beta(z)$ from Eq.~\eqref{WardId_gamma_beta_z} and then $\alpha(z)$ from Eq.~\eqref{Def_gamma_z}. After a long but straightforward calculation using systematically, in order to remove higher-order derivatives, the differential relation \cite{AS65,Olver:2010:NHMF,NIST:DLMF}
\begin{equation}
\label{DER_HypegeometricFn}
\frac{d}{dz} F(a,b;c;z) = \frac{ab}{c} F(a+1,b+1;c+1;z)
\end{equation}
as well as the hypergeometric differential equation \eqref{Def_HypegeometricDiffEq}, we obtain
\begin{widetext}
\allowdisplaybreaks
\begin{align}
\allowdisplaybreaks
\label{Sol_WVectorFn}
G^{(+)A}_{\mu \nu'}(x,x') = &\Big[
- (2/3) C^1_\gamma \, z(1-z) \, F'(5/2+\lambda, 5/2-\lambda; 3; z)
+ (2/3) C^2_\gamma \, z(1-z) \, F'(5/2+\lambda, 5/2-\lambda; 3; 1-z)
\nonumber \\ &\hphantom{\Big[}
- C^1_\gamma \, (1-2z) \, F(5/2+\lambda, 5/2-\lambda; 3; z)
- C^2_\gamma \, (1-2z) \, F(5/2+\lambda, 5/2-\lambda; 3; 1-z)
\nonumber \\ &\hphantom{\Big[}
- (1/4) C^1_G \, F(5/2+\kappa, 5/2-\kappa; 3; z)
+ (1/4) C^2_G \, F(5/2+\kappa, 5/2-\kappa; 3; 1-z)
\Big] \, g_{\mu \nu'}
\nonumber \\
+ \frac{8}{R} \, &\Big[
- C^1_\gamma \, F'(5/2+\lambda, 5/2-\lambda; 3; z)
+ C^2_\gamma \, F'(5/2+\lambda, 5/2-\lambda; 3; 1-z)
\nonumber \\ &\hphantom{\Big[}
- 3 C^1_\gamma \, (1/z) \, F(5/2+\lambda, 5/2-\lambda; 3; z)
- 3 C^2_\gamma \, (1/z) \, F(5/2+\lambda, 5/2-\lambda; 3; 1-z)
\nonumber \\ &\hphantom{\Big[}
- (3/4) C^1_G \, F'(5/2+\kappa, 5/2-\kappa; 3; z)
- (3/4) C^2_G \, F'(5/2+\kappa, 5/2-\kappa; 3; 1-z)
\nonumber \\ &\hphantom{\Big\{}
- (3/4) C^1_G \, (1/z) \, F(5/2+\kappa, 5/2-\kappa; 3; z)
+ (3/4) C^2_G \, (1/z) \, F(5/2+\kappa, 5/2-\kappa; 3; 1-z)
\Big] \, z_{;\mu} z_{;\nu'}.
\end{align}
\end{widetext}
It is important to note that, in Eq.~\eqref{Sol_WVectorFn} as well in the following of the article, the derivative of the hypergeometric function $F(a,b;c;z)$ with respect to its argument $z$ is denoted by $F'(a,b;c;z)$.

In summary, the general form of the Wightman function associated with the massive vector field $A_\mu$ is explicitly given by Eq.~\eqref{Sol_WVectorFn} while the Wightman function associated with the scalar field $\Phi$ and the ghost fields $C$ and $C^\ast$ is explicitly given by Eq.~\eqref{Sol_WScalarFn}. In the following subsections, we shall fix the integration constants $C^1_G$, $C^2_G$, $C^1_\gamma$ and $C^2_\gamma$ appearing in the expression of these two-point functions.

\subsubsection{Wightman functions for Hadamard vacua}
\label{Sec.IIc2}

Previously, we have assumed that the vacuum $|0 \rangle $ of Stueckelberg electromagnetism is of Hadamard type. Here, we shall consider that this assumption can be realized by imposing that, at short distance, i.e. for $x' \to x$, the Wightman function (\ref{Wightman_A}) associated with the massive vector field $A_\mu$ and the Wightman function (\ref{WGreenFn}) associated with both the scalar field $\Phi $ and the ghost fields $C$ and $C^\ast$ satisfy
\begin{align}
\label{WA_HadamardSingularity}
& g^{\mu \nu'} G^{(+)A}_{\mu \nu'}(x,x') \underset{x' \to x}{\thicksim} \frac{1}{2\pi^2} \, \frac{1}{\sigma(x,x')}
\end{align}
and
\begin{align}
\label{Wscalar_HadamardSingularity}
& G^{(+)}(x,x') \underset{x' \to x}{\thicksim} \frac{1}{8\pi^2} \, \frac{1}{\sigma(x,x')}.
\end{align}
It should be noted that, at first sight, the conditions (\ref{WA_HadamardSingularity}) and (\ref{Wscalar_HadamardSingularity}) are less constraining than assuming that the Feynman propagators associated with all the fields of the theory can be represented in the Hadamard form \cite{Belokogne:2015etf}. (Without loss of generality, in this discussion, we focus on Feynman propagators but it would be possible to consider, equivalently, Hadamard Green functions.) Indeed, this last assumption provides stronger constraints on the geometrical coefficients of the singular terms in $1/[\sigma (x,x') + i\epsilon] $ and $\ln [\sigma (x,x') + i\epsilon] $ of the Hadamard representations of the Feynman propagators. In fact, due to the choice $\xi = 1$ of the gauge parameter, we know that the two-point functions of the Stueckelberg theory can be represented in the Hadamard form and, as a consequence, if we fix the dominant term of the coefficient of $1/[\sigma (x,x') + i\epsilon] $, all the other terms are unambiguously determined [see the differential equation (A3a) and the boundary condition (A3b) as well as the recursion relations (39a), (39b), (43a) and (43b) in Ref.~\cite{Belokogne:2015etf}].

By inserting \eqref{Def_z_Ra} or \eqref{Def_z_Rb} into \eqref{Sol_WVectorFn} and \eqref{Sol_WScalarFn}, we obtain the short distance expansions
\begin{widetext}
\begin{align}
\allowdisplaybreaks
\label{Expansion_WVectorFn}
& g^{\mu \nu'} G^{(+)A}_{\mu \nu'}(x,x') =
\left[
- \frac{ 3072 \, C^1_\gamma}{\Gamma(5/2+\lambda)\Gamma(5/2-\lambda)}
- \frac{2304 \, C^1_G}{\Gamma(5/2+\kappa)\Gamma(5/2-\kappa)}
\right]  \, \frac{1}{R \sigma^2(x,x')}
\nonumber \\ & \quad
+ \left[
- \frac{32 \, (85/4 - \lambda^2) \, C^1_\gamma}{\Gamma(5/2+\lambda)\Gamma(5/2-\lambda)}
- \frac{72 \, (19/4 + \kappa^2) \, C^1_G}{\Gamma(5/2+\kappa)\Gamma(5/2-\kappa)}
\right] \, \frac{1}{R \sigma (x,x')} + \underset{x' \to x}{O}[\ln \sigma (x,x')]
\end{align}
\end{widetext}
and
\begin{align}
\label{Expansion_WScalarFn}
& G^{(+)}(x,x') = \frac{24 \, C^1_G}{\Gamma(3/2+\kappa)\Gamma(3/2-\kappa)}  \, \frac{1}{R \sigma (x,x')} \nonumber \\ & \quad + \underset{x' \to x}{O}[\ln \sigma (x,x')]
\end{align}
and, by comparing with \eqref{WA_HadamardSingularity} and \eqref{Wscalar_HadamardSingularity}, we can fix the two integration constants $C^1_\gamma$ and $C^1_G$. We have
\begin{subequations}
\allowdisplaybreaks
\label{Const_C1gamma}
\begin{align}
C^1_\gamma &= - \frac{R}{256\pi^2} \, \frac{\Gamma(5/2+\lambda) \Gamma(5/2-\lambda)}{1/4-\lambda^2}
 \label{Const_C1gamma_a} \\ &=
- \frac{R}{256\pi} \, \frac{9/4-\lambda^2}{\cos(\pi \lambda)} \label{Const_C1gamma_b}
\end{align}
\end{subequations}
and
\begin{subequations}
\allowdisplaybreaks
\label{Const_C1G}
\begin{align}
C^1_G &= \frac{R}{192\pi^2} \, \Gamma(3/2+\kappa) \Gamma(3/2-\kappa) \label{Const_C1G_a}
 \\ &=
  \frac{R}{192\pi} \, \frac{1/4-\kappa^2}{\cos(\pi \kappa)}. \label{Const_C1G_b}
\end{align}
\end{subequations}
Here, in order simplify the expressions \eqref{Const_C1gamma_a} and \eqref{Const_C1G_a} which involve the Gamma function $\Gamma (z)$, we have used the reflection formula \cite{AS65,Olver:2010:NHMF,NIST:DLMF}
\begin{align}
\label{GammaFunction_ref}
\Gamma (z) \Gamma (1-z)= \frac{\pi}{\sin (\pi z)}.
\end{align}

\subsubsection{Wightman functions in $\mathrm {dS}^4$}
\label{Sec.IIc3}

In $\mathrm {dS}^4$, in order to fix the remaining integration constants $C^2_\gamma$ and $C^2_G$, we require the regularity of the Wightman functions \eqref{Sol_WVectorFn} and \eqref{Sol_WScalarFn} at the antipodal point of $x$, $x'=Px$ and therefore on its light cone), or, in other words, for $z \to 0$ (see also Fig.~\ref{fig_dS_PenroseDiagram}) \cite{Schomblond:1976xc,Allen:1985ux}. We obtain immediately
\begin{subequations}
\allowdisplaybreaks
\label{Const_C2_dS}
\begin{align}
\label{Const_C2gamma_dS}
C^2_\gamma = 0
\end{align}
and
\begin{align}
\label{Const_C2G_dS}
C^2_G = 0.
\end{align}
\end{subequations}

By inserting now the integration constants \eqref{Const_C1gamma}, \eqref{Const_C1G}, \eqref{Const_C2gamma_dS} and \eqref{Const_C2G_dS} into the general expressions \eqref{Sol_WVectorFn} and \eqref{Sol_WScalarFn}, we obtain in $\mathrm {dS}^4$ the explicit expressions of the Wightman function associated with the massive vector field $A_\mu$ and the Wightman function associated with both the Stueckelberg scalar field $\Phi$ and the ghost fields $C$ and $C^\ast$. We have
\begin{widetext}
\allowdisplaybreaks
\begin{align}
\allowdisplaybreaks
\label{Sol_WVectorFn_dS_a}
G^{(+)A}_{\mu \nu'}(x,x') =  \frac{H^2}{32\pi} &\bigg[
  \frac{9/4-\lambda^2}{\cos(\pi \lambda)} \, z(1-z) \, F'(5/2+\lambda, 5/2-\lambda; 3; z)
+ \frac{3}{2} \, \frac{(9/4-\lambda^2)}{\cos(\pi \lambda)} \, (1-2z) \, F(5/2+\lambda, 5/2-\lambda; 3; z)
\nonumber \\ &\hphantom{\bigg[}
- \frac{1}{2} \, \frac{(1/4-\kappa^2)}{\cos(\pi \kappa)} \, F(5/2+\kappa, 5/2-\kappa; 3; z)
\bigg] \, g_{\mu \nu'}
\nonumber \\
+ \frac{1}{32\pi} &\bigg[
  \frac{9/4-\lambda^2}{\cos(\pi \lambda)} \, F'(5/2+\lambda, 5/2-\lambda; 3; z)
+ 3 \, \frac{(9/4-\lambda^2)}{\cos(\pi \lambda)} \, (1/z) \, F(5/2+\lambda, 5/2-\lambda; 3; z)
\nonumber \\ &\hphantom{\bigg[}
- \frac{1/4-\kappa^2}{\cos(\pi \kappa)} \, F'(5/2+\kappa, 5/2-\kappa; 3; z)
- \frac{1/4-\kappa^2}{\cos(\pi \kappa)} \, (1/z) \, F(5/2+\kappa, 5/2-\kappa; 3; z)
\bigg] \, z_{;\mu} z_{;\nu'}
\end{align}
and
\begin{align}
\label{Sol_WScalarFn_dS_a}
G^{(+)}(x,x') =  \frac{H^2}{16\pi} \, \frac{1/4-\kappa^2}{\cos(\pi \kappa)} \, F(3/2+\kappa, 3/2-\kappa; 2; z).
\end{align}

\end{widetext}
It should be noted that \eqref{Sol_WVectorFn_dS_a} is in accordance with the result obtained recently by Fr{\"o}b and Higuchi from a mode-sum construction [see Eq.~(25) in Ref.~\cite{Frob:2013qsa}] while \eqref{Sol_WScalarFn_dS_a} is a closed form which can be found in various works concerning the massive scalar field in $\mathrm {dS}^4$ \cite{Dowker:1975tf,Schomblond:1976xc,Bunch:1978yq,Allen:1985ux,Allen:1985wd}.

\subsubsection{Wightman functions in $\mathrm {AdS}^4$}
\label{Sec.IIc4}

In $\mathrm {AdS}^4$, in order to fix the remaining integration constants $C^2_\gamma$ and $C^2_G$, we require that the Wightman functions \eqref{Sol_WVectorFn} and \eqref{Sol_WScalarFn} fall off as fast as possible at spatial infinity, i.e., for $z \to \infty$ (see also Fig.~\ref{fig_AdS_PenroseDiagram}). We recall that this condition is imposed because the Cauchy problem is not well posed in $\mathrm {AdS}^4$, this gravitational background being not globally hyperbolic \cite{HawkingEllis}. Such a condition permits us to control the flow of information through spatial infinity \cite{Avis:1977yn,Breitenlohner:1982jf}.

In the expressions \eqref{Sol_WVectorFn} and \eqref{Sol_WScalarFn} of the Wightman functions, the hypergeometric functions are expressed in term of the variables $z$ and $1-z$. In order to impose the boundary condition previously mentioned, it is helpful to reexpress them in term of the variable $1/z$. This can be achieved thanks to the connection formulas \cite{AS65,Olver:2010:NHMF,NIST:DLMF}
\begin{widetext}
\allowdisplaybreaks
\begin{subequations}
\allowdisplaybreaks
\label{NIST_ConnectionFormula_25_27}
\begin{align}
\allowdisplaybreaks
\label{NIST_ConnectionFormula_25}
F(a,b;c;z) =
  \frac{\Gamma(c)\Gamma(b-a)}{\Gamma(b)\Gamma(c-a)} \, (-z)^{-a} F(a,a-c+1;a-b+1;1/z)
+ \frac{\Gamma(c)\Gamma(a-b)}{\Gamma(a)\Gamma(c-b)} \, (-z)^{-b} F(b,b-c+1;b-a+1;1/z)
\end{align}
which is valid for $|\mathrm{arg} (-z)| < \pi$ and
\begin{align}
\allowdisplaybreaks
\label{NIST_ConnectionFormula_27}
F(a,b;c;1-z) =
  \frac{\Gamma(c)\Gamma(b-a)}{\Gamma(b)\Gamma(c-a)} \, z^{-a} F(a,c-b;a-b+1;1/z)
+ \frac{\Gamma(c)\Gamma(a-b)}{\Gamma(a)\Gamma(c-b)} \, z^{-b} F(b,c-a;b-a+1;1/z)
\end{align}
\end{subequations}
\end{widetext}
which is valid for $|\mathrm{arg} (z)| < \pi$. We can then observe that the Wightman functions \eqref{Sol_WVectorFn} and \eqref{Sol_WScalarFn} approach zero as fast as possible at spatial infinity if we eliminate the terms in $z^{-(5/2-\kappa)}$ and $z^{-(5/2-\lambda)}$ in the expression of the former and the term in $z^{-(3/2-\kappa)}$ in the expression of the latter. (Here, since $m >0$, we have that $\lambda = \sqrt{1/4 + m^2/K^2} >0$ and $\kappa = \sqrt{9/4 + m^2/K^2} >0$.) We then obtain immediately
\begin{subequations}
\allowdisplaybreaks
\label{Const_C2_AdS}
\begin{align}
\label{Const_C2gamma_AdS}
C^2_\gamma = \pm \, i \, e^{\pm i\pi \lambda} C^1_\gamma
\end{align}
and
\begin{align}
\label{Const_C2G_AdS}
C^2_G = \mp \, i \, e^{\pm i\pi \kappa} C^1_G.
\end{align}
\end{subequations}
In Eqs.~\eqref{Const_C2gamma_AdS} and \eqref{Const_C2G_AdS}, the upper sign (the lower sign) must be chosen if, in the expressions \eqref{Sol_WVectorFn} and \eqref{Sol_WScalarFn} of the Wightman functions, $z$ lies in the upper half plane (in the lower half plane). This follows from the relation $(-z)^{-a} = e^{\mp i\pi a} z^{-a}$ which is a consequence of        $(-z)^{-a}=\exp [- a \ln(-z)]$.

By inserting now the integration constants \eqref{Const_C1gamma}, \eqref{Const_C1G}, \eqref{Const_C2gamma_AdS} and \eqref{Const_C2G_AdS} into the general expressions \eqref{Sol_WVectorFn} and \eqref{Sol_WScalarFn}, we can obtain in $\mathrm {AdS}^4$ the explicit expressions of the Wightman function associated with the massive vector field $A_\mu$ and the Wightman function associated with both the Stueckelberg scalar field $\Phi$ and the ghost fields $C$ and $C^\ast$. Making use of the reflection formula \eqref{GammaFunction_ref} and of the duplication formula \cite{AS65,Olver:2010:NHMF,NIST:DLMF}
\begin{align}
\label{GammaFunction_dup}
\Gamma (2 z) = \frac{2^{2z}}{2 \sqrt{\pi}} \, \Gamma (z) \Gamma (z +1/2)
\end{align}
to deal with the $\Gamma$-function, we have
\begin{widetext}
\allowdisplaybreaks
\begin{subequations}
\label{Sol_WVectorFn_AdS_ab}
\allowdisplaybreaks
\begin{align}
\allowdisplaybreaks
\label{Sol_WVectorFn_AdS_a}
G^{(+)A}_{\mu \nu'}(x,x') = - \frac{K^2}{32\pi} &\bigg\{
  \frac{9/4-\lambda^2}{\cos(\pi \lambda)} \, z(1-z) \, \left[ F'(5/2+\lambda, 5/2-\lambda; 3; z) \mp \, i \, e^{\pm i\pi\lambda} F'(5/2+\lambda, 5/2-\lambda; 3; 1-z) \right]
\nonumber \\ &\hphantom{\bigg\{}
+ \frac{3}{2} \, \frac{(9/4-\lambda^2)}{\cos(\pi \lambda)} \, (1-2z) \, \left[ F(5/2+\lambda, 5/2-\lambda; 3; z) \pm \, i \, e^{\pm i\pi\lambda} F(5/2+\lambda, 5/2-\lambda; 3; 1-z) \right]
\nonumber \\ &\hphantom{\bigg\{}
- \frac{1}{2} \, \frac{(1/4-\kappa^2)}{\cos(\pi \kappa)} \, \left[ F(5/2+\kappa, 5/2-\kappa; 3; z) \pm \, i \, e^{\pm i\pi\kappa} F(5/2+\kappa, 5/2-\kappa; 3; 1-z) \right]
\bigg\} \, g_{\mu \nu'}
\nonumber \\
+ \frac{1}{32\pi} &\bigg\{
  \frac{9/4-\lambda^2}{\cos(\pi \lambda)} \, \left[ F'(5/2+\lambda, 5/2-\lambda; 3; z) \mp \, i \, e^{\pm i\pi\lambda} F'(5/2+\lambda, 5/2-\lambda; 3; 1-z) \right]
\nonumber \\ &\hphantom{\bigg\{}
+ 3 \, \frac{(9/4-\lambda^2)}{\cos(\pi \lambda)} \, (1/z) \, \left[ F(5/2+\lambda, 5/2-\lambda; 3; z) \pm \, i \, e^{\pm i\pi\lambda} F(5/2+\lambda, 5/2-\lambda; 3; 1-z) \right]
\nonumber \\ &\hphantom{\bigg\{}
- \frac{1/4-\kappa^2}{\cos(\pi \kappa)} \, \left[ F'(5/2+\kappa, 5/2-\kappa; 3; z) \mp \, i \, e^{\pm i\pi\kappa} F'(5/2+\kappa, 5/2-\kappa; 3; 1-z) \right]
\nonumber \\ &\hphantom{\bigg\{}
- \frac{1/4-\kappa^2}{\cos(\pi \kappa)} \, (1/z) \, \left[ F(5/2+\kappa, 5/2-\kappa; 3; z) \pm \, i\, e^{\pm i\pi\kappa} F(5/2+\kappa, 5/2-\kappa; 3; 1-z) \right]
\bigg\} \, z_{;\mu} z_{;\nu'}
\\
\label{Sol_WVectorFn_AdS_b}
= - \frac{K^4}{32\pi^2m^2} &\bigg\{
  \frac{\Gamma(7/2+\lambda)\Gamma(1/2+\lambda)}{\Gamma(1+2\lambda)} \, z^{-(5/2+\lambda)} (1-z) \, F(7/2+\lambda, 1/2+\lambda; 1+2\lambda; 1/z)
\nonumber \\ &\hphantom{\bigg\{}
- 3 \, \frac{\Gamma(5/2+\lambda)\Gamma(1/2+\lambda)}{\Gamma(1+2\lambda)} \, z^{-(5/2+\lambda)} (1-2z) \, F(5/2+\lambda, 1/2+\lambda; 1+2\lambda; 1/z)
\nonumber \\ &\hphantom{\bigg\{}
+ \frac{\Gamma(5/2+\kappa)\Gamma(1/2+\kappa)}{\Gamma(1+2\kappa)} \, z^{-(5/2+\kappa)} \, F(5/2+\kappa, 1/2+\kappa; 1+2\kappa; 1/z)
\bigg\} \, g_{\mu \nu'}
\nonumber \\
+ \frac{K^2}{32\pi^2m^2} &\bigg\{
  \frac{\Gamma(7/2+\lambda)\Gamma(1/2+\lambda)}{\Gamma(1+2\lambda)} \, z^{-(7/2+\lambda)} \, F(7/2+\lambda, 1/2+\lambda; 1+2\lambda; 1/z)
\nonumber \\ &\hphantom{\bigg\{}
- 6 \, \frac{\Gamma(5/2+\lambda)\Gamma(1/2+\lambda)}{\Gamma(1+2\lambda)} \, z^{-(7/2+\lambda)} \, F(5/2+\lambda, 1/2+\lambda; 1+2\lambda; 1/z)
\nonumber \\ &\hphantom{\bigg\{}
- \frac{\Gamma(7/2+\kappa)\Gamma(1/2+\kappa)}{\Gamma(1+2\kappa)} \, z^{-(7/2+\kappa)} \, F(7/2+\kappa, 1/2+\kappa; 1+2\kappa; 1/z)
\nonumber \\ &\hphantom{\bigg\{}
+ 2 \, \frac{\Gamma(5/2+\kappa)\Gamma(1/2+\kappa)}{\Gamma(1+2\kappa)} \, z^{-(7/2+\kappa)} \, F(5/2+\kappa, 1/2+\kappa; 1+2\kappa; 1/z)
\bigg\} \, z_{;\mu} z_{;\nu'}
\end{align}
\end{subequations}
and
\begin{subequations}
\label{Sol_WScalarFn_AdS_ab}
\allowdisplaybreaks
\begin{align}
\label{Sol_WScalarFn_AdS_a}
G^{(+)}(x,x') &= - \frac{K^2}{16\pi} \, \frac{1/4-\kappa^2}{\cos(\pi\kappa)} \, \left[ F(3/2+\kappa, 3/2-\kappa; 2; z) \mp \, i \, e^{ \pm i\pi \kappa} F(3/2+\kappa, 3/2-\kappa; 2; 1-z) \right]
\\
\label{Sol_WScalarFn_AdS_b}
&= \frac{K^2}{16\pi^2} \, \frac{\Gamma(3/2+\kappa)\Gamma(1/2+\kappa)}{\Gamma(1+2\kappa)} \, z^{-(3/2+\kappa)} \, F(3/2+\kappa, 1/2+\kappa; 1+2\kappa; 1/z).
\end{align}
\end{subequations}
\end{widetext}
Here, it is important to recall that, in Eqs.~\eqref{Sol_WVectorFn_AdS_a} and \eqref{Sol_WScalarFn_AdS_a}, the upper sign (the lower sign) must be chosen if $z$ lies in the upper half plane (in the lower half plane). It should be noted that we have provided two equivalent expressions for these two-point functions: the hypergeometric functions appearing in formulas \eqref{Sol_WVectorFn_AdS_a} and \eqref{Sol_WScalarFn_AdS_a} are given in term of the variables $z$ and $1-z$ while those appearing in formulas \eqref{Sol_WVectorFn_AdS_b} and \eqref{Sol_WScalarFn_AdS_b} are expressed in term of $1/z$. The expression (\ref{Sol_WScalarFn_AdS_ab}) is a result which can be found in some other works concerning the massive scalar field in $\mathrm {AdS}^4$ (see, e.g., Refs.~\cite{Burgess:1984ti,Allen:1985wd,Camporesi:1992wn,Kent:2014nya}). To our knowledge, the Wightman function \eqref{Sol_WVectorFn_AdS_ab} associated with the massive vector field $A_\mu$ of the Stueckelberg theory is not in the literature. In Ref.~\cite{Janssen:1986fz}, Janssen and Dullemond have considered that problem but we are unable to link their results with ours.

\subsection{Feynman propagators and Hadamard Green functions in $\mathrm {dS}^4$ and $\mathrm {AdS}^4$}
\label{Sec.IId}

It is well known that, in quantum field theory in flat spacetime, we can construct all the interesting two-point functions, i.e., the retarded and advanced Green functions, the Feynman propagator and the Hadamard Green function, from the Wightman function by taking its real or its imaginary part and, when it is necessary, by using multiplication by a step function in time. In some sense, this remains true in curved spacetime \cite{DeWitt:1960fc,Birrell:1982ix,Fulling:1989nb} and, in this subsection, we shall provide the expressions of the Feynman propagators and the Hadamard Green functions of Stueckelberg massive electromagnetism from the Wightman functions obtained previously. Here, we shall adopt a pragmatic point of view by following the approach and the arguments of Allen and Jacobson \cite{,Allen:1985ux,Allen:1985wd}. It is however interesting to note the existence of a more rigorous point of view exposed in impressive articles by Bros, Epstein and Moschella which concern the scalar field in de Sitter spacetime \cite{Bros:1995js} and general quantum field theories in anti-de Sitter spacetime \cite{Bros:2001tk}.

\subsubsection{In $\mathrm {dS}^4$}
\label{Sec.IId1}

The expressions \eqref{Sol_WVectorFn_dS_a} and \eqref{Sol_WScalarFn_dS_a} of the Wightman functions involve the hypergeometric function $F(a,b;c;z)$ which, in general, has a branch point at $z=1$ (i.e., for $x'$ on the light cone of $x$) and a branch cut which runs along the real axis from $z=1$ to $+\infty$ (i.e., for $x'$ in the light cone of $x$) [see also Eq.~\eqref{Range_z_dS}]. As a consequence, these Wightman functions are perfectly defined when $x$ and $x'$ are spacelike related or cannot be joined by a geodesic but, when they are timelike related, it is important to specify how to approach the branch cut. In fact, it is necessary to replace in Eqs.~\eqref{Sol_WVectorFn_dS_a} and \eqref{Sol_WScalarFn_dS_a} the biscalar $z$ by the biscalar $z \mp i \epsilon$ (here $\epsilon \to 0_+$) where the minus sign (respectively the plus sign) is chosen when $x'$ lies in the past (respectively the future) of $x$. Indeed, in $\mathrm {dS}^4$, due to the relation \eqref{Def_z} [note that $z=1 - (R/24) \sigma + \dots $], the prescription $z \to z \mp i \epsilon$ induces the change $\sigma \to \sigma \pm i \epsilon$ which permits us to encode the usual behavior of the Wightman functions in curved spacetime (see also Chap.~4 of Ref.~\cite{Wald:1995yp}) and to recover, in the flat-space limit, the Wightman functions of Minkowski quantum field theory.

The Feynman propagator $G^{A}_{\mu \nu'} (x,x') = i \langle 0 | T A_\mu(x) A_{\nu'} (x') | 0 \rangle $ associated with the massive vector field $A_\mu$ (here, $T$ denotes the time-ordering operator) is obtained from the Wightman function \eqref{Sol_WVectorFn_dS_a} by writing $G^{A}_{\mu \nu'} (x,x')=i \, G^{(+)A}_{\mu \nu'} (z - i \epsilon)$ with $\epsilon \to 0_+$. Indeed, in $\mathrm {dS}^4$, the prescription $z \to z - i \epsilon$ induces the change $\sigma \to \sigma + i \epsilon$ which permits us to encode the time ordering (see also Secs.~II~C and III~A of Ref.~\cite{Belokogne:2015etf}). Similarly, the Feynman propagators $G^{\Phi} (x,x') = i \langle \psi | T \Phi(x) \Phi(x') | \psi \rangle$ and $G^\mathrm{Gh} (x,x') = i \langle \psi | T C^\ast(x) C(x') | \psi \rangle $ associated respectively with the scalar field $\Phi$ and the ghost fields $C$ and $C^\ast$ are equals to $i \, G^{(+)}(z - i \epsilon)$.

The expression of the Hadamard Green function
\begin{align}
\label{G1_A}
G^{(1)A}_{\mu \nu'} (x,x') = \langle 0 |  A_\mu (x)  A_{\nu'} (x') + A_{\nu'} (x')A_\mu (x)  | 0 \rangle
\end{align}
associated with the massive vector field $A_\mu$ is obviously obtained from \eqref{Sol_WVectorFn_dS_a} by noting that
\begin{align}
\label{Sol_G1VectorFn}
G^{(1)A}_{\mu \nu'} (x,x')=G^{(+)A}_{\mu \nu'} (x,x') + G^{(+)A}_{\nu' \mu} (x',x).
\end{align}
In the same way, the Hadamard Green function
\begin{align}
\label{G1_Phi}
G^{(1)\Phi}(x,x') = \langle 0 |  \Phi (x) \Phi (x') + \Phi (x') \Phi (x)  | 0 \rangle
\end{align}
associated with the auxiliary scalar field $\Phi$ and the Hadamard Green function
\begin{align}
\label{G1_Gh}
G^{(1)\mathrm{Gh}}(x,x') = \langle 0 |  C^\ast (x) C (x') - C (x') C^\ast (x)  | 0 \rangle
\end{align}
associated with the ghost fields $C$ and $C^\ast$ can be obtained from \eqref{Sol_WScalarFn_dS_a}. We have
\begin{align}
\label{G1_Scalar_PhiGh}
G^{(1)\Phi}(x,x')=G^{(1)\mathrm{Gh}}(x,x')=G^{(1)}(x,x')
\end{align}
with
\begin{align}
\label{Sol_G1ScalarFn}
G^{(1)}(x,x') = G^{(+)}(x,x')+G^{(+)}(x',x).
\end{align}
Formulas \eqref{Sol_G1VectorFn} and \eqref{Sol_G1ScalarFn} must be taken with a grain of salt. Indeed, it is important to recall that Hadamard Green functions are real-valued while Wightman functions are complex-valued and the prescription permitting us to define the former and the latter on the branch cut are different. In fact, Hadamard Green functions are average accross the cut, i.e., we have
\begin{align}
\label{Sol_G1VectorFn_average}
G^{(1)A}_{\mu \nu'} (x,x')=G^{(+)A}_{\mu \nu'} (z + i \epsilon) + G^{(+)A}_{\mu \nu'} (z - i \epsilon)
\end{align}
and
\begin{align}
\label{Sol_G1ScalarFn_average}
G^{(1)}(x,x') = G^{(+)}(z+ i \epsilon)+G^{(+)}(z - i \epsilon)
\end{align}
with $\epsilon \to 0_+$. This prescription, which is in total agreement with that defining the Wightman functions, permits us to have at our disposal Hadamard Green functions which are real-valued. More explicitly, by inserting \eqref{Sol_WVectorFn_dS_a} into \eqref{Sol_G1VectorFn_average} and \eqref{Sol_WScalarFn_dS_a} into \eqref{Sol_G1ScalarFn_average}, we obtain
\begin{widetext}
\allowdisplaybreaks
\begin{align}
\allowdisplaybreaks
\label{Sol_G1VectorFn_dS_a}
G^{(1)A}_{\mu \nu'}(x,x') =  \frac{H^2}{16\pi} &\bigg[
  \frac{9/4-\lambda^2}{\cos(\pi \lambda)} \, z(1-z) \, (\mathrm{Re}F)'(5/2+\lambda, 5/2-\lambda; 3; z)
+ \frac{3}{2} \, \frac{(9/4-\lambda^2)}{\cos(\pi \lambda)} \, (1-2z) \, (\mathrm{Re}F)(5/2+\lambda, 5/2-\lambda; 3; z)
\nonumber \\ &\hphantom{\bigg[}
- \frac{1}{2} \, \frac{(1/4-\kappa^2)}{\cos(\pi \kappa)} \, (\mathrm{Re}F)(5/2+\kappa, 5/2-\kappa; 3; z)
\bigg] \, g_{\mu \nu'}
\nonumber \\
+ \frac{1}{16\pi} &\bigg[
  \frac{9/4-\lambda^2}{\cos(\pi \lambda)} \, (\mathrm{Re}F)'(5/2+\lambda, 5/2-\lambda; 3; z)
+ 3 \, \frac{(9/4-\lambda^2)}{\cos(\pi \lambda)} \, (1/z) \, (\mathrm{Re}F)(5/2+\lambda, 5/2-\lambda; 3; z)
\nonumber \\ &\hphantom{\bigg[}
- \frac{1/4-\kappa^2}{\cos(\pi \kappa)} \, (\mathrm{Re}F)'(5/2+\kappa, 5/2-\kappa; 3; z)
- \frac{1/4-\kappa^2}{\cos(\pi \kappa)} \, (1/z) \, (\mathrm{Re}F)(5/2+\kappa, 5/2-\kappa; 3; z)
\bigg] \, z_{;\mu} z_{;\nu'}.
\end{align}
\end{widetext}
and
\begin{align}
\label{Sol_G1ScalarFn_dS_a}
&G^{(1)}(x,x') =  \frac{H^2}{8\pi} \, \frac{1/4-\kappa^2}{\cos(\pi \kappa)} \, (\mathrm{Re}F)(3/2+\kappa, 3/2-\kappa; 2; z).
\end{align}
Here, we have introduced the average accross the cut of the hypergeometric function $F(a,b;c;z)$ defined by
\begin{subequations}
\label{Re_and_Im_FHyper}
\begin{align}
\label{Re_FHyper}
(\mathrm{Re}F)(a,b;c;z) = \frac{F(a,b;c;z+i \epsilon)+F(a,b;c;z-i \epsilon)}{2}.
\end{align}
This function is nothing else than the real part of $F(a,b;c;z)$ on the branch cut. We have also used its derivative $(\mathrm{Re}F)'(a,b;c;z)$ with respect to its argument $z$. It should be noted that below we shall need also its imaginary part
\begin{align}
\label{Im_FHyper}
(\mathrm{Im}F)(a,b;c;z) = \frac{F(a,b;c;z+i \epsilon)-F(a,b;c;z-i \epsilon)}{2i}
\end{align}
and we shall use its derivative $(\mathrm{Im}F)'(a,b;c;z)$ with respect to its argument $z$.
\end{subequations}

\subsubsection{In $\mathrm {AdS}^4$}
\label{Sec.IId2}

\textit{Mutatis mutandis}, the previous discussion can be adapted to obtain the Feynman propagators and the Hadamard Green functions in $\mathrm {AdS}^4$. We first note that the branch cut of the Wightman functions \eqref{Sol_WVectorFn_AdS_ab} and \eqref{Sol_WScalarFn_AdS_ab} runs along the real axis from $z=-\infty$ to $z=1$. This appears clearly if we consider the expressions \eqref{Sol_WVectorFn_AdS_b} and \eqref{Sol_WScalarFn_AdS_b}. Indeed, the functions of the form $z^{-a}=\exp (-a \ln z)$ and the hypergeometric functions of the form $F(a,b;c;1/z)$ involved in these expressions are respectively cut along the negative axis and the segment $[0,1]$. As a consequence, the Wightman functions \eqref{Sol_WVectorFn_AdS_ab} and \eqref{Sol_WScalarFn_AdS_ab} are perfectly defined if $z>1$, i.e., when $x$ and $x'$ are spacelike related [see also Eq.~\eqref{Range_z_AdS}] whereas, if $z \le 1$, and in particular when $x$ and $x'$ are timelike related (i.e., if $z \in ]0,1[$), it is important to specify how to approach the branch cut. In fact, it is necessary to replace in Eqs.~\eqref{Sol_WVectorFn_AdS_ab} and \eqref{Sol_WScalarFn_AdS_ab} the biscalar $z$ by the biscalar $z \pm i \epsilon$ (here $\epsilon \to 0_+$) where the plus sign (respectively the minus sign) is chosen when $x'$ lies in the past (respectively the future) of $x$. Indeed, in $\mathrm {AdS}^4$, due to the relation \eqref{Def_z}, it is now the prescription $z \to z \pm i \epsilon$ which induces the change $\sigma \to \sigma \pm i \epsilon$ permitting us to encode the usual behavior of the Wightman functions in curved spacetime.

In $\mathrm {AdS}^4$, the Feynman propagator $G^{A}_{\mu \nu'} (x,x') = i \langle 0 | T A_\mu(x) A_{\nu'} (x') | 0 \rangle $ associated with the massive vector field $A_\mu$ is obtained from the Wightman function \eqref{Sol_WVectorFn_AdS_ab} by writing $G^{A}_{\mu \nu'} (x,x')=i \, G^{(+)A}_{\mu \nu'} (z + i \epsilon)$ with $\epsilon \to 0_+$. Indeed, in this gravitational background, the prescription $z \to z + i \epsilon$ induces the change $\sigma \to \sigma + i \epsilon$ which permits us to encode the time ordering. Similarly, the Feynman propagators $G^{\Phi} (x,x') = i \langle \psi | T \Phi(x) \Phi(x') | \psi \rangle$ and $G^\mathrm{Gh} (x,x') = i \langle \psi | T C^\ast(x) C(x') | \psi \rangle $ associated respectively with the scalar field $\Phi$ and the ghost fields $C$ and $C^\ast$ are equals to $i \, G^{(+)}(z + i \epsilon)$.

In $\mathrm {AdS}^4$, the expression of the Hadamard Green function \eqref{G1_A} associated with the massive vector field $A_\mu$ is obtained by inserting \eqref{Sol_WVectorFn_AdS_ab} into \eqref{Sol_G1VectorFn_average} while the Hadamard Green function \eqref{G1_Scalar_PhiGh} associated with both the massive scalar field $\Phi$ and the ghost fields $C$ and $C^\ast$ is obtained by inserting \eqref{Sol_WScalarFn_AdS_ab} into \eqref{Sol_G1ScalarFn_average}. In fact, here we shall construct the Hadamard Green functions from \eqref{Sol_WVectorFn_AdS_a} and \eqref{Sol_WScalarFn_AdS_a} only. Indeed, the resulting expressions are easily tractable in the context of the renormalization of the stress-energy tensor, or more precisely, their regular parts can be naturally extracted. Of course, in order to obtain the Hadamard Green functions, it is then important to take carefully into account the upper sign (in that case, we use the variable $z + i \epsilon$ and we are working in the upper half plane) or the lower sign (in that case, we use the variable $z - i \epsilon$ and we are working in the lower half plane) in \eqref{Sol_WVectorFn_AdS_a} and \eqref{Sol_WScalarFn_AdS_a}. A straightforward calculation leads to expressions which are explicitly real-valued and given by
\begin{widetext}
\allowdisplaybreaks
\begin{align}
\allowdisplaybreaks
\label{Sol_G1VectorFn_AdS_a}
G^{(1)A}_{\mu \nu'}(x,x') = - \frac{K^2}{16\pi} &\bigg\{
  \frac{9/4-\lambda^2}{\cos(\pi \lambda)} \, z(1-z) \, \left[ (\mathrm{Re}F)'(5/2+\lambda, 5/2-\lambda; 3; z) + \sin (\pi\lambda) (\mathrm{Re}F)'(5/2+\lambda, 5/2-\lambda; 3; 1-z) \right.  \nonumber  \\
  & \left. \qquad\qquad\qquad\qquad    - \cos (\pi\lambda) (\mathrm{Im}F)'(5/2+\lambda, 5/2-\lambda; 3; 1-z) \right]
\nonumber \\ &\hphantom{\bigg\{}
+ \frac{3}{2} \, \frac{(9/4-\lambda^2)}{\cos(\pi \lambda)} \, (1-2z) \, \left[ (\mathrm{Re}F)(5/2+\lambda, 5/2-\lambda; 3; z) - \sin (\pi\lambda) (\mathrm{Re}F)(5/2+\lambda, 5/2-\lambda; 3; 1-z) \right.  \nonumber  \\
  & \left. \qquad\qquad\qquad\qquad  + \cos (\pi\lambda) (\mathrm{Im}F)(5/2+\lambda, 5/2-\lambda; 3; 1-z) \right]
\nonumber \\ &\hphantom{\bigg\{}
- \frac{1}{2} \, \frac{(1/4-\kappa^2)}{\cos(\pi \kappa)} \, \left[ (\mathrm{Re}F)(5/2+\kappa, 5/2-\kappa; 3; z) - \sin (\pi\kappa) (\mathrm{Re}F)(5/2+\kappa, 5/2-\kappa; 3; 1-z) \right.  \nonumber  \\
  & \left. \qquad\qquad\qquad\qquad + \cos (\pi\kappa) (\mathrm{Im}F)(5/2+\kappa, 5/2-\kappa; 3; 1-z) \right]
\bigg\} \, g_{\mu \nu'}
\nonumber \\
+ \frac{1}{16\pi} &\bigg\{
  \frac{9/4-\lambda^2}{\cos(\pi \lambda)} \, \left[ (\mathrm{Re}F)'(5/2+\lambda, 5/2-\lambda; 3; z) + \sin (\pi\lambda) (\mathrm{Re}F)'(5/2+\lambda, 5/2-\lambda; 3; 1-z) \right.  \nonumber  \\
  & \left. \qquad\qquad\qquad\qquad - \cos (\pi\lambda) (\mathrm{Im}F)'(5/2+\lambda, 5/2-\lambda; 3; 1-z) \right]
\nonumber \\ &\hphantom{\bigg\{}
+ 3 \, \frac{(9/4-\lambda^2)}{\cos(\pi \lambda)} \, (1/z) \, \left[ (\mathrm{Re}F)(5/2+\lambda, 5/2-\lambda; 3; z)  - \sin (\pi\lambda) (\mathrm{Re}F)(5/2+\lambda, 5/2-\lambda; 3; 1-z) \right.  \nonumber  \\
  & \left. \qquad\qquad\qquad\qquad + \cos (\pi\lambda) (\mathrm{Im}F)(5/2+\lambda, 5/2-\lambda; 3; 1-z) \right]
\nonumber \\ &\hphantom{\bigg\{}
- \frac{1/4-\kappa^2}{\cos(\pi \kappa)} \, \left[ (\mathrm{Re}F)'(5/2+\kappa, 5/2-\kappa; 3; z) + \sin (\pi\kappa) (\mathrm{Re}F)'(5/2+\kappa, 5/2-\kappa; 3; 1-z) \right.  \nonumber  \\
  & \left. \qquad\qquad\qquad\qquad - \cos (\pi\kappa) (\mathrm{Im}F)'(5/2+\kappa, 5/2-\kappa; 3; 1-z) \right]
\nonumber \\ &\hphantom{\bigg\{}
- \frac{1/4-\kappa^2}{\cos(\pi \kappa)} \, (1/z) \, \left[ (\mathrm{Re}F)(5/2+\kappa, 5/2-\kappa; 3; z) - \sin (\pi\kappa) (\mathrm{Re}F)(5/2+\kappa, 5/2-\kappa; 3; 1-z) \right.  \nonumber  \\
  & \left. \qquad\qquad\qquad\qquad + \cos (\pi\kappa) (\mathrm{Im}F)(5/2+\kappa, 5/2-\kappa; 3; 1-z) \right]
\bigg\} \, z_{;\mu} z_{;\nu'}
\end{align}
and
\begin{align}
\label{Sol_G1ScalarFn_AdS_a}
G^{(1)}(x,x') &= - \frac{K^2}{8\pi} \, \frac{1/4-\kappa^2}{\cos(\pi\kappa)} \, \left[ (\mathrm{Re}F)(3/2+\kappa, 3/2-\kappa; 2; z) + \sin (\pi\kappa) (\mathrm{Re}F)(3/2+\kappa, 3/2-\kappa; 2; 1-z) \right.  \nonumber \\
& \qquad\qquad\qquad\qquad \left. - \cos (\pi\kappa) (\mathrm{Im}F)(3/2+\kappa, 3/2-\kappa; 2; 1-z) \right].
\end{align}
\end{widetext}

\section{Renormalized stress-energy tensor of Stueckelberg electromagnetism}
\label{Sec.III}

In this section, from the general formalism developed in Ref.~\cite{Belokogne:2015etf}, we shall obtain exact analytical expressions for the vacuum expectation value of the RSET of the massive vector field propagating in $\mathrm {dS}^4$ and $\mathrm {AdS}^4$. We shall, in particular, fix the geometrical ambiguities in the results (see Refs.~\cite{Wald:1978pj,Tichy:1998ws} for interesting remarks on the ambiguity problem as well as Sec.~IV~E of Ref.~\cite{Belokogne:2015etf} and references therein) by considering the flat-space limit and, moreover, we shall discuss the zero-mass limit of the expressions found.

\subsection{General considerations}
\label{Sec.IIIa}

In this subsection, we have gathered some results established in Ref.~\cite{Belokogne:2015etf} which will be necessary to construct, in the next three subsections, the RSETs associated with Stueckelberg electromagnetism in $\mathrm {dS}^4$ and $\mathrm {AdS}^4$. By doing so, we hope to alleviate the task of the reader and to prevent him from drowning in the heavy formalism developed in our previous article.

We have seen in Ref.~\cite{Belokogne:2015etf} that, in the context of the renormalization of the stress-energy tensor of Stueckelberg electromagnetism, it is necessary to extract the regular and state-dependent parts of the Hadamard Green functions $G^{(1)A}_{\mu \nu'} (x,x')$ and $G^{(1)}(x,x')$. They can be obtained by removing from $G^{(1)A}_{\mu \nu'} (x,x')$ and $G^{(1)}(x,x')$ their singular and purely geometrical parts $G^{(1)A}_\mathrm{sing}{}_{\mu \nu'} (x,x')$ and $G^{(1)}_\mathrm{sing} (x,x')$. We have
\begin{align}
\label{HadamardRep_G1Vect_reg}
G^{(1)A}_\mathrm{reg}{}_{\mu \nu'} (x,x') =  G^{(1)A}_{\mu \nu'} (x,x') - G^{(1)A}_\mathrm{sing}{}_{\mu \nu'} (x,x') \end{align}
and
\begin{align}
\label{HadamardRep_G1Scal_reg}
G^{(1)}_\mathrm{reg} (x,x')  = G^{(1)} (x,x') - G^{(1)}_\mathrm{sing} (x,x')
\end{align}
where
\begin{eqnarray}
\label{HadamardRep_G1Vect_sing}
&& G^{(1)A}_\mathrm{sing}{}_{\mu \nu'} (x,x') = \frac{1}{4\pi^2} \, \left[  \frac{\Delta^{1/2}(x,x')}{\sigma(x,x')} \, g_{\mu \nu'}(x,x')
\right. \nonumber \\ && \quad \left. \vphantom{\frac{\Delta^{1/2}(x,x')}{\sigma(x,x')}}
+ V^A_{\mu \nu'}(x,x') \ln \left| M^2 \,\sigma(x,x') \right|
\right]
\end{eqnarray}
and
\begin{eqnarray}
\label{HadamardRep_G1Scal_sing}
&& G^{(1)}_\mathrm{sing} (x,x') = \frac{1}{4\pi^2} \, \left[
\frac{\Delta^{1/2}(x,x')}{\sigma(x,x')}
\right. \nonumber \\ && \quad \left. \vphantom{\frac{\Delta^{1/2}(x,x')}{\sigma(x,x')}}
+ V(x,x') \ln \left| M^2 \sigma(x,x') \right|
\right].
\end{eqnarray}
The expressions of $G^{(1)A}_\mathrm{sing}{}_{\mu \nu'} (x,x')$ and $G^{(1)}_\mathrm{sing} (x,x')$ involve the geodetic interval $\sigma(x,x')$, the bivector of parallel transport $g_{\mu \nu'}(x,x')$, the Van Vleck-Morette determinant $\Delta (x,x')$ (see Ref.~\cite{DeWitt:1960fc} or the Appendix of Ref.~\cite{Belokogne:2015etf} for its definition and properties) as well as the geometrical bivector $V^A_{\mu \nu'}(x,x')$ and the geometrical biscalar $V(x,x')$ which are defined by the expansions $V^A_{\mu \nu'}(x,x') = \sum_{n=0}^{+\infty} V^A_n{}_{\mu \nu'}(x,x') \, \sigma^n(x,x')$ and $V(x,x') = \sum_{n=0}^{+\infty} V_n (x,x') \, \sigma^n(x,x') $ and by the recursion relations satisfied by the coefficients $V^A_n{}_{\mu \nu'}(x,x')$ and $V_n(x,x')$ [see Eqs.~(39a), (39b), (43a) and (43b) in Ref.~\cite{Belokogne:2015etf}]. Moreover, we have introduced the renormalization mass $M$ permitting us to make dimensionless the argument of the logarithm.

In fact, in order to construct the RSET, we only need the lower coefficients of the covariant Taylor series expansions for $x' \to x$ of the bitensor
\begin{align}
\label{WVect_def}
W^A_{\mu \nu}(x,x') = 4 \pi^2 \, g_{\nu}^{\phantom{\nu} \nu'}(x,x') G^{(1)A}_\mathrm{reg}{}_{\mu \nu'} (x,x')
\end{align}
and of the biscalar
\begin{align}
\label{WScal_def}
W (x,x') = 4 \pi^2 \, G^{(1)}_\mathrm{reg}(x,x')
\end{align}
or, more precisely, we only need the covariant Taylor series expansions of these two quantities up to order $\sigma^1 (x,x')$ [see Sec.~IV of Ref.~\cite{Belokogne:2015etf}]. As a consequence, we are not really interested by the full expressions of \eqref{HadamardRep_G1Vect_sing} and \eqref{HadamardRep_G1Scal_sing} but by their covariant Taylor series expansions truncated by neglecting the terms vanishing faster than $\sigma (x,x')$ for $x' \to x$. They can be obtained from the covariant Taylor series expansion of $\Delta^{1/2}(x,x')$ up to order $\sigma^2(x,x')$ [see Eq.~(A9) in Ref.~\cite{Belokogne:2015etf}], the covariant Taylor series expansion of $V^A_{\mu \nu}(x,x')$ up to order $\sigma^1(x,x')$ [see Eqs.~(71a)--(72d) in Ref.~\cite{Belokogne:2015etf}] and the covariant Taylor series expansion of $V(x,x')$ up to order $\sigma^1(x,x')$ [see Eqs.~(73a)--(74c) in Ref.~\cite{Belokogne:2015etf}]. By inserting these covariant Taylor series expansions into \eqref{HadamardRep_G1Vect_sing} and \eqref{HadamardRep_G1Scal_sing}, we can derive the covariant Taylor series expansions of $W^A_{\mu \nu}(x,x')$ defined by \eqref{WVect_def} [see Eq.~(75) in Ref.~\cite{Belokogne:2015etf} for its general expression] and of $W (x,x')$ defined by \eqref{WScal_def} [see Eq.~(76) in Ref.~\cite{Belokogne:2015etf} for its general expression].

Fortunately, in maximally symmetric spacetimes, due to the relations \eqref{Expr_RiemannTensor}--\eqref{Expr_RicciScalar}, the regularization of the Hadamard Green functions $G^{(1)A}_{\mu \nu'} (x,x')$ and $G^{(1)}(x,x')$ greatly simplifies. Indeed :
\begin{enumerate}[label=(\roman*)]

\item The covariant Taylor series expansions of the various bitensors involved in the singular parts of the Hadamard Green functions reduce to
    \begin{subequations}
\begin{align}
\label{CTS_DetVVM}
& \Delta^{1/2} = 1 + (1/24) \, R  \sigma \nonumber \\
& \qquad + (19/17280) \, R^2  \sigma^2 + O( \sigma^3 ),
\end{align}
\begin{align}
\label{CTS_VVector}
& V^A_{\mu \nu} =  \left\{ \left[(1/2) \, m^2 + (1/24) \, R \right]  + \left[(1/8) \, m^4 \right.\right. \nonumber \\
& \quad \left.\left. + (1/24) \,m^2 R + (1/432) \, R^2 \right] \sigma \right\} g_{\mu \nu}   \nonumber \\
& \quad +(1/1728) R^2 \, \sigma_{;\mu}\sigma_{;\nu} + O( \sigma^{3/2} ),
\end{align}
\noindent and
\begin{align}
\label{CTS_VScalar}
& V =   \left[ (1/2) \, m^2 - (1/12) \, R \right] \nonumber \\
& \qquad + \left[ (1/8) \, m^4 - (1/48) \, m^2 R \right]  \sigma  \nonumber \\
& \qquad + O( \sigma^{3/2} ).
\end{align}
\end{subequations}
\item The covariant Taylor series expansions of the bitensors $W^A_{\mu \nu}(x,x')$ and $W (x,x')$ reduce to
\begin{eqnarray}
\label{covTaylorSeries_WVector}
& W^A_{\mu \nu} = s_{\mu \nu} + \frac{1}{2} \,  s_{\mu \nu a b} \sigma^{;a} \sigma^{;b}
+ O( \sigma^{3/2} )
\end{eqnarray}
and
\begin{align}
\label{covTaylorSeries_WScalar}
& W = w  + \frac{1}{2} \, w_{a b} \, \sigma^{;a} \sigma^{;b} + O( \sigma^{3/2} ) .
\end{align}

\end{enumerate}
\noindent In the following, these considerations will facilitate our task.

In our previous article, we have provided two different expressions for the RSET of Stueckelberg electromagnetism:

\begin{enumerate}[label=(\roman*)]

\item A first expression which only involves state-dependent as well as geometrical quantities associated with the massive vector field $A_\mu$ [see Eq.~(123) in Ref.~\cite{Belokogne:2015etf}]. Here, the contribution of the quantum massive scalar field $\Phi$ has been removed thanks to a Ward identity and the result obtained is given in terms of the first coefficients of the covariant Taylor series expansions for $x' \to x$ of the bitensor $W^A_{\mu \nu}(x,x')$ [see Eq.~(75) in Ref.~\cite{Belokogne:2015etf}].

\item A second expression where the contributions of the massive vector field $A_\mu$ and of the massive scalar field $\Phi$ have been artificially separated [see Eqs.~(125) and (126) in Ref.~\cite{Belokogne:2015etf}] and which has been constructed in such a way that the zero-mass limit of the first contribution reduces to the RSET of Maxwell's electromagnetism. The result obtained is given in terms of the first coefficients of the covariant Taylor series expansions for $x' \to x$ of the bitensors $W^A_{\mu \nu}(x,x')$ and $W (x,x')$ [see Eqs.~(75) and (76) in Ref.~\cite{Belokogne:2015etf}].

\end{enumerate}
\noindent Of course, these expressions are equivalent but it will be necessary to consider both in order to clearly discuss the zero-mass limit of the Stueckelberg theory (see Sec.~\ref{Sec.IIId}). In maximally symmetric spacetimes, these expressions simplify considerably. Indeed, the RSET of Stueckelberg electromagnetism can be expressed in term of its trace as
\begin{equation}
\label{SET_MaximallySym}
\langle 0 | \widehat{T}_{\mu \nu} | 0 \rangle_\mathrm{ren} = \frac{1}{4} \, \langle 0 | \widehat{T}^{\phantom{\rho} \rho}_{\rho} | 0 \rangle_\mathrm{ren} \, g_{\mu \nu}
\end{equation}
and this one reduces to
\begin{align}
\label{SET_MaximallySym_TR1}
& \langle 0 | \widehat{T}^{\phantom{\rho} \rho}_{\rho} | 0 \rangle_\mathrm{ren} = \frac{1}{8\pi^2}
\left\{
\left[- m^2 -(1/8) \, R \right] s^{\phantom{\rho} \rho}_{\rho}
\right. \nonumber \\ & \left. \qquad
+ s^{\phantom{\rho \tau} \rho \tau}_{\rho \tau}
+ 3 \, v^A_1{}^{\phantom{\rho} \rho}_{\rho}
\right\}
+ \Theta^{\phantom{\rho} \rho}_{\rho}
\end{align}
if we focus on the expression which only involves the characteristics of the quantum massive vector field $A_\mu$ [see, in Ref.~\cite{Belokogne:2015etf}, Eq.~(123) as well as Eq.~(124)]. If we alternatively focus on the expression which involves both the characteristics of the massive vector field $A_\mu$ and of the massive scalar field $\Phi$ [see, in Ref.~\cite{Belokogne:2015etf}, Eqs.~(125) and (126) as well as Eqs.~(127) and (128)], we have
\begin{equation}
\label{SET_MaximallySym_TR2}
\langle 0 | \widehat{T}^{\phantom{\rho} \rho}_{\rho} | 0 \rangle_\mathrm{ren} = \mathcal{T}^A{}^{\phantom{\rho} \rho}_{\rho} + \mathcal{T}^\Phi{}^{\phantom{\rho} \rho}_{\rho} + \Theta^{\phantom{\rho} \rho}_{\rho}
\end{equation}
\noindent with
\begin{subequations}
\label{SET_MaximallySym_TR2_A_Phi}
\begin{eqnarray}
\label{SET_MaximallySym_TR2_A}
&& \mathcal{T}^A{}^{\phantom{\rho} \rho}_{\rho} = \frac{1}{8\pi^2} \left\{
\left[- m^2 -(1/4) \, R \right] s^{\phantom{\rho} \rho}_{\rho} + 2 \, s^{\phantom{\rho \tau} \rho \tau}_{\rho \tau}
+ 4 \, v^A_1{}^{\phantom{\rho} \rho}_{\rho}
\right\} \nonumber \\
&&
\end{eqnarray}
and
\begin{equation}
\label{SET_MaximallySym_TR2_Phi}
 \mathcal{T}^\Phi{}^{\phantom{\rho} \rho}_{\rho} = \frac{1}{8\pi^2} \left(  - m^2 w + 2 \, v_1 \right).
\end{equation}
\end{subequations}
It should be noted that the term $\Theta^{\phantom{\rho} \rho}_{\rho}$ appearing in Eqs.~\eqref{SET_MaximallySym_TR1} and \eqref{SET_MaximallySym_TR2} is a purely geometrical term which encodes the ambiguities in the definition of the RSET [see Sec.~IV~E of Ref.~\cite{Belokogne:2015etf}] and which involves, in particular, a contribution associated with the renormalization mass $M$. We also recall that the term $v^A_1{}^{\phantom{\rho} \rho}_{\rho}$ in Eqs.~\eqref{SET_MaximallySym_TR1} and \eqref{SET_MaximallySym_TR2_A} and the term $v_1$ in Eq.~\eqref{SET_MaximallySym_TR2_Phi} are purely geometrical quantities which appear in the covariant Taylor series expansions of $V^A_{\mu \nu}(x,x')$ and $V (x,x')$ [see Eqs.~(71b) and (73b) in Ref.~\cite{Belokogne:2015etf}]. From Eqs.~(72d) and (74c) of Ref.~\cite{Belokogne:2015etf}, we can show that, in maximally symmetric backgrounds, they reduce to
\begin{equation}
\label{TRv1A_MaximallySym}
v^A_1{}^{\phantom{\rho} \rho}_{\rho} = (1/2) \, m^4 + (1/12) \, m^2 R - (1/2160) \, R^2
\end{equation}
and
\begin{equation}
\label{v1_MaximallySym}
v_1 = (1/8) \, m^4 - (1/24) \, m^2 R + (29/8640) \, R^2.
\end{equation}

It is crucial to discuss the form of the trace term $\Theta^{\phantom{\rho} \rho}_{\rho}$ appearing in Eqs.~\eqref{SET_MaximallySym_TR1} and \eqref{SET_MaximallySym_TR2}. In an arbitrary gravitational background, such a term is given by Eq.~(133) of Ref.~\cite{Belokogne:2015etf} which reduces, in maximally symmetric spacetimes, to
\begin{equation}
\label{AmbGEN_MaximallySym}
\Theta^{\phantom{\rho} \rho}_{\rho} = \frac{1}{8\pi^2} \left( \alpha \, m^4 + \beta \, m^2 R \right).
\end{equation}
Here, $\alpha$ and $\beta$ are constants which can be fixed by imposing additional physical conditions on the RSET (see below) or which can be redefined by renormalization of Newton's gravitational constant and of the cosmological constant. Indeed, let us recall that, in Eq.~\eqref{AmbGEN_MaximallySym}, the terms $\alpha \, m^4$ and $\beta \, m^2 R$ come from the Einstein-Hilbert action defining the dynamics of the gravitational field (see also the discussion in Sec.~IV~E~1 of Ref.~\cite{Belokogne:2015etf}). Of course, the ambiguities associated with the renormalization mass $M$ are necessarily of this form but their expressions are totally determined. In maximally symmetric spacetimes, the renormalization mass induces in \eqref{SET_MaximallySym_TR1} a contribution given by
\begin{equation}
\label{AmbGEN_MaximallySym_TR1}
\Theta^{\phantom{\rho} \rho}_{\rho} (M) = \frac{\ln(M^2)}{8\pi^2} \left[ (3/2) \, m^4 + (1/4) \, m^2 R  \right].
\end{equation}
It can be derived from Eq.~(146) of Ref.~\cite{Belokogne:2015etf} which is valid in an arbitrary gravitational background. Similarly, the renormalization mass induces in \eqref{SET_MaximallySym_TR2} a contribution given by
\begin{equation}
\label{AmbGEN_MaximallySym_TR2_Sum}
\Theta^{\phantom{\rho} \rho}_{\rho} (M) = \Theta^A{}^{\phantom{\rho} \rho}_{\rho} (M) + \Theta^\Phi{}^{\phantom{\rho} \rho}_{\rho} (M)
\end{equation}
with
\begin{subequations}
\label{AmbGEN_MaximallySym_TR2}
\begin{equation}
\label{AmbGEN_MaximallySym_TR2_A}
\Theta^A{}^{\phantom{\rho} \rho}_{\rho} (M) = \frac{\ln(M^2)}{8\pi^2} \left[  m^4 + (1/3) \, m^2 R  \right]
\end{equation}
and
\begin{equation}
\label{AmbGEN_MaximallySym_TR2_Phi}
\Theta^\Phi{}^{\phantom{\rho} \rho}_{\rho} (M) = \frac{\ln(M^2)}{8\pi^2} \left[  (1/2) \, m^4 - (1/12) \, m^2 R  \right].
\end{equation}
\end{subequations}
It can be derived from Eqs.~(144), (147a) and (147b) of Ref.~\cite{Belokogne:2015etf} which are valid in an arbitrary gravitational background.

To conclude this subsection, we would like to remark that the Taylor coefficient $s^{\phantom{\rho \tau} \rho \tau}_{\rho \tau}$ appearing in the expressions \eqref{SET_MaximallySym_TR1} and \eqref{SET_MaximallySym_TR2_A} of the RSET can be related with lower-order Taylor coefficients. Indeed,
due to the ``Ward identity'' linking the bitensors $W^A_{\mu \nu}(x,x')$ and $W(x,x')$ [see Eq.~(85) in Ref.~\cite{Belokogne:2015etf}], we can write in a maximally symmetric spacetime [see Eq.~(86b) in Ref.~\cite{Belokogne:2015etf}]
\begin{align}
\label{Relation_coefTaylorSeries_WA_W_2}
s^{\phantom{\rho \tau} \rho \tau}_{\rho \tau}  =  (1/8) \, R s^{\phantom{\rho} \rho}_\rho + w^{\phantom{\rho} \rho}_\rho - v^A_1{}^{\phantom{\rho} \rho}_{\rho} +4 \, v_1 .
\end{align}
Moreover, due to the ``wave equation'' satisfied by the biscalar $W(x,x')$ [see Eq.~(82) in Ref.~\cite{Belokogne:2015etf}], we have the constraint [see Eq.~(83a) in Ref.~\cite{Belokogne:2015etf}]
\begin{equation}
\label{Relation_coefTaylorSeries_W_1}
w^{\phantom{\rho} \rho}_\rho = m^2 \, w  - 6 \, v_1.
\end{equation}
By inserting \eqref{Relation_coefTaylorSeries_W_1} into \eqref{Relation_coefTaylorSeries_WA_W_2}, we then obtain
\begin{align}
\label{Relation_coefTaylorSeries_WA_W}
s^{\phantom{\rho \tau} \rho \tau}_{\rho \tau}  =   (1/8) \, R s^{\phantom{\rho} \rho}_\rho + m^2 \, w  - v^A_1{}^{\phantom{\rho} \rho}_{\rho} - 2 \, v_1.
\end{align}
This last equation is very interesting. Indeed, it permits us to realize that, in maximally symmetric spacetimes, the construction of the RSET of Stueckelberg electromagnetism can be accomplished using only the coefficients $s^{\phantom{\rho} \rho}_\rho$ and $w$, i.e., the lowest-order coefficients of the covariant Taylor series expansions \eqref{covTaylorSeries_WVector} and \eqref{covTaylorSeries_WScalar}. As a consequence, in the regularization process, it would be sufficient to consider the covariant Taylor series expansions of \eqref{HadamardRep_G1Vect_sing} and \eqref{HadamardRep_G1Scal_sing} truncated by neglecting the terms vanishing faster than $\sigma^0 (x,x')=1$ for $x' \to x$. Moreover, we can remark that the Taylor coefficient $s^{\phantom{\rho} \rho}_\rho$ is a trace term. So, in order to obtain its expression, it would be sufficient to regularize the trace $g^{\mu \nu'} (x,x')G^{(1)A}_{\mu \nu'} (x,x')$ of the Hadamard Green functions \eqref{Sol_G1VectorFn_dS_a} or \eqref{Sol_G1VectorFn_AdS_a} and this could be done rather easily by taking into account the relation \eqref{GeometricalRelations_a}. In fact, in the following, we shall not use these considerations even if it is obvious they could greatly simplify our job. We intend to determine the full covariant Taylor series expansions \eqref{covTaylorSeries_WVector} and \eqref{covTaylorSeries_WScalar} because this will permit us to control the internal consistency of our calculations and, in particular, that the singular terms in $\ln |\sigma (x,x')|$ hidden in the expressions \eqref{Sol_G1VectorFn_dS_a}, \eqref{Sol_G1ScalarFn_dS_a}, \eqref{Sol_G1VectorFn_AdS_a} and \eqref{Sol_G1ScalarFn_AdS_a} of the Hadamard Green functions are of Hadamard type. We shall use the constraints \eqref{Relation_coefTaylorSeries_W_1} and \eqref{Relation_coefTaylorSeries_WA_W} only to check our results.

\subsection{The renormalized stress-energy tensor in $\mathrm {dS}^4$}
\label{Sec.IIIb}

In $\mathrm {dS}^4$, the covariant Taylor series expansions \eqref{covTaylorSeries_WVector} and \eqref{covTaylorSeries_WScalar} can be obtained by inserting the expressions \eqref{Sol_G1VectorFn_dS_a} and \eqref{Sol_G1ScalarFn_dS_a} of the Hadamard Green functions into \eqref{HadamardRep_G1Vect_reg} and \eqref{HadamardRep_G1Scal_reg} taking into account (i) the relation \eqref{Def_z_H} which links the quadratic form $z(x,x')$ with the geodetic interval $\sigma(x,x')$ as well as (ii) the expression of the singular parts \eqref{HadamardRep_G1Vect_sing} and \eqref{HadamardRep_G1Scal_sing} constructed by using the covariant Taylor series expansions \eqref{CTS_DetVVM}, \eqref{CTS_VVector} and \eqref{CTS_VScalar}. We then obtain
\begin{widetext}
\begin{subequations}
\label{TaylorCoeff_in_dS_WAetW}
\begin{eqnarray}
&&  s_{\mu \nu} = \Big\{ - (1/2) \, m^2 - (19/144) \, R
   + \big[ (3/8) \, m^2 + (1/16) \, R \big] \big[ \ln(R/24) + \Psi(5/2+\lambda) + \Psi(5/2-\lambda) + 2 \gamma \big] \nonumber \\
&&  \hphantom{s_{\mu \nu} = }
   + \big[ (1/8) \, m^2 - (1/48) \, R \big] \big[ \ln(R/24) + \Psi(5/2+\kappa) + \Psi(5/2-\kappa) + 2 \gamma \big]
\Big\} \, g_{\mu \nu} , \label{TaylorCoeff_in_dS_WA_a}\\
&&  s_{\mu \nu a b} = \Big\{ - (5/16) \, m^4 - (43/288) \, m^2 R - (691/51840) \, R^2 + \nonumber \\
&&  \hphantom{s_{\mu \nu a b} = }
   + \big[ (5/48) \, m^4 + (5/72) \, m^2 R + (5/576) \, R^2 \big] \big[ \ln(R/24) + \Psi(7/2+\lambda) + \Psi(7/2-\lambda) + 2 \gamma \big] \nonumber \\
&&  \hphantom{s_{\mu \nu a b} = }
   - \big[ (1/32) \, m^2 R + (1/192) \, R^2 \big] \big[ \ln(R/24) + \Psi(5/2+\lambda) + \Psi(5/2-\lambda) + 2 \gamma \big] \nonumber \\
&&  \hphantom{s_{\mu \nu a b} = }
   + \big[ (1/48) \, m^4 + (1/288) \, m^2 R - (1/864) \, R^2 \big] \big[ \ln(R/24) + \Psi(7/2+\kappa) + \Psi(7/2-\kappa) + 2 \gamma \big]
\Big\} \, g_{\mu \nu} g_{a b} \nonumber \\
&&  \hphantom{s_{\mu \nu a b} } + \Big\{ (1/144) \, m^2 R + (19/5184) \, R^2 \nonumber \\
&&  \hphantom{s_{\mu \nu a b} = }
   - \big[ (1/24) \, m^4 + (1/36) \, m^2 R + (1/288) \, R^2 \big] \big[ \ln(R/24) + \Psi(7/2+\lambda) + \Psi(7/2-\lambda) + 2 \gamma \big] \nonumber \\
&&  \hphantom{s_{\mu \nu a b} = }
   + \big[ (1/32) \, m^2 R + (1/192) \, R^2 \big] \big[ \ln(R/24) + \Psi(5/2+\lambda) + \Psi(5/2-\lambda) + 2 \gamma \big] \nonumber \\
&&  \hphantom{s_{\mu \nu a b} = }
   + \big[ (1/24) \, m^4 + (1/144) \, m^2 R - (1/432) \, R^2 \big] \big[ \ln(R/24) + \Psi(7/2+\kappa) + \Psi(7/2-\kappa) + 2 \gamma \big] \nonumber \\
&&  \hphantom{s_{\mu \nu a b} = }
   - \big[ (1/96) \, m^2 R - (1/576) \, R^2 \big] \big[ \ln(R/24) + \Psi(5/2+\kappa) + \Psi(5/2-\kappa) + 2 \gamma \big]
\Big\} \, g_{\mu (a|} g_{\nu |b)} \label{TaylorCoeff_in_dS_WA_b}
\end{eqnarray}
and
\begin{eqnarray}
&&  w = - (1/2) \, m^2 + (1/18) \, R + \big[ (1/2) \, m^2 - (1/12) \, R \big] \big[ \ln(R/24) + \Psi(3/2+\kappa) + \Psi(3/2-\kappa) + 2 \gamma \big], \label{TaylorCoeff_in_dS_W_c}\\
&&  w_{a b} = \Big\{ - (5/16) \, m^4 + (13/288) \, m^2 R + (1/5760) \, R^2 \nonumber \\
&&  \hphantom{w_{a b} = } + \big[ (1/8) \, m^4 - (1/48) \, m^2 R \big] \big[ \ln(R/24) + \Psi(5/2+\kappa) + \Psi(5/2-\kappa) + 2 \gamma \big] \Big\} \, g_{a b} . \label{TaylorCoeff_in_dS W_d}
\end{eqnarray}
\end{subequations}
In the previous expressions, we have introduced the Digamma function $\Psi(z)=(d/dz) \ln \Gamma (z)$ and we have used systematically its properties \cite{AS65,Olver:2010:NHMF,NIST:DLMF} and, more particularly, the recurrence formula
\begin{equation}
\label{Psi_z_ppté1}
\Psi(z+1)= \Psi(z) + 1/z
\end{equation}
in order to simplify them. We have also introduced the Euler-Mascheroni constant $\gamma=-\Psi(1)$. From these results, we can now write
\begin{subequations}
\label{Tr_TaylorCoeff_in_dS_WAetW}
\begin{eqnarray}
&&  s^{\phantom{\rho} \rho}_\rho = - 2 \, m^2 - (19/36) \, R
   + \big[ (3/2) \, m^2 + (1/4) \, R \big] \big[ \ln(R/24) + \Psi(5/2+\lambda) + \Psi(5/2-\lambda) + 2 \gamma \big] \nonumber \\
&&  \hphantom{s_{\mu \nu} = }
   + \big[ (1/2) \, m^2 - (1/12) \, R \big] \big[ \ln(R/24) + \Psi(5/2+\kappa) + \Psi(5/2-\kappa) + 2 \gamma \big] , \label{Tr_TaylorCoeff_in_dS_WA_a}\\
&&  s^{\phantom{\rho \tau} \rho \tau}_{\rho \tau}  =  - (5/4) \, m^4 - (19/36) \, m^2 R - (1/60) \, R^2  \nonumber \\
&&  \hphantom{s^{\phantom{\rho \tau} \rho \tau}_{\rho \tau} = }
   + \big[ (3/16) \, m^2 R + (1/32) \, R^2 \big] \big[ \ln(R/24) + \Psi(5/2+\lambda) + \Psi(5/2-\lambda) + 2 \gamma \big] \nonumber \\
&&  \hphantom{ s^{\phantom{\rho \tau} \rho \tau}_{\rho \tau} = }
   + \big[ (1/2) \, m^4 + (1/12) \, m^2 R - (1/36) \, R^2 \big] \big[ \ln(R/24) + \Psi(7/2+\kappa) + \Psi(7/2-\kappa) + 2 \gamma \big] \nonumber \\
&&  \hphantom{ s^{\phantom{\rho \tau} \rho \tau}_{\rho \tau} = }
    -\big[ (5/48) \, m^2 R - (5/288) \, R^2 \big] \big[ \ln(R/24) + \Psi(5/2+\kappa) + \Psi(5/2-\kappa) + 2 \gamma \big]\label{Tr_TaylorCoeff_in_dS_WA_b}
\end{eqnarray}
and
\begin{eqnarray}
&&  w^{\phantom{\rho} \rho}_\rho =  - (5/4) \, m^4 + (13/72) \, m^2 R + (1/1440) \, R^2 \nonumber\\
&& \qquad\quad +  \big[ (1/2) \, m^4 - (1/12) \, m^2 R \big] \big[ \ln(R/24) + \Psi(5/2+\kappa) + \Psi(5/2-\kappa) + 2 \gamma \big] . \label{Tr_TaylorCoeff_in_dS_W_c}
\end{eqnarray}
\end{subequations}

\end{widetext}
It should be noted that the constraints \eqref{Relation_coefTaylorSeries_W_1} and \eqref{Relation_coefTaylorSeries_WA_W} are satisfied by these coefficients. This can be easily checked using the recurrence formula \eqref{Psi_z_ppté1}.

By inserting now the expressions \eqref{Tr_TaylorCoeff_in_dS_WA_a} and \eqref{Tr_TaylorCoeff_in_dS_WA_b} into \eqref{SET_MaximallySym_TR1} taking into account the geometrical term \eqref{TRv1A_MaximallySym} as well as the geometrical ambiguities \eqref{AmbGEN_MaximallySym} and \eqref{AmbGEN_MaximallySym_TR1}, we have for the trace of the RSET

\begin{widetext}
\begin{eqnarray}
\label{SET_dS4_TR1}
&& 8\pi^2 \langle 0 | \widehat{T}^{\phantom{\rho} \rho}_{\rho} | 0 \rangle_\mathrm{ren \,\,\mathrm {dS}^4} = (\alpha + 9/4) \, m^4 + (\beta + 17/24) m^2 R + (19/1440) \, R^2
\nonumber \\
&& \phantom{8\pi^2 \langle 0 | \widehat{T}^{\phantom{\rho} \rho}_{\rho} | 0 \rangle_\mathrm{ren \,\,\mathrm {dS}^4} = } - \big[ (3/2) \, m^4 + (1/4) \, m^2 R \big] \big[ \ln(R/(24 M^2)) + \Psi(5/2+\lambda) + \Psi(5/2-\lambda) + 2 \gamma \big]
\end{eqnarray}
\end{widetext}
and, of course, the RSET $\langle 0 | \widehat{T}_{\mu \nu} | 0 \rangle_\mathrm{ren}$ can be obtained immediately from \eqref{SET_MaximallySym}. This expression could be considered as the final result of our work in $\mathrm {dS}^4$. Indeed, by construction, it fully includes the state-dependence of the Stueckelberg theory and, moreover, it takes into account all the geometrical ambiguities. However, it is possible to go further and to fix the renormalization mass $M$ and the coefficient $\alpha$ by requiring the vanishing of this expression in the flat-space limit, i.e. for $R \to 0$. We first absorb the term $2 \gamma$ into the renormalization mass $M$ and we then obtain $M=m/\sqrt{2}$ and $\alpha = -9/4$ which leads to
\begin{widetext}
\begin{eqnarray}
\label{SET_dS4_TR1_def}
&& 8\pi^2 \langle 0 | \widehat{T}^{\phantom{\rho} \rho}_{\rho} | 0 \rangle_\mathrm{ren \,\,\mathrm {dS}^4} = (\beta + 17/24) \, m^2 R + (19/1440) \, R^2
\nonumber \\
&& \phantom{8\pi^2 \langle 0 | \widehat{T}^{\phantom{\rho} \rho}_{\rho} | 0 \rangle_\mathrm{ren \,\,\mathrm {dS}^4} = } - \big[ (3/2) \, m^4 + (1/4) \, m^2 R \big] \big[ \ln(R/(12 m^2)) + \Psi(5/2+\lambda) + \Psi(5/2-\lambda) \big] .
\end{eqnarray}
\end{widetext}
This last result is not free of ambiguities due to the arbitrary coefficient $\beta$ remaining in the expression of the trace of the RSET. However, it is worth noting that it would be possible to cancel it or, more precisely, to cancel the term $(\beta + 17/24) m^2 R$ by a finite renormalization of the Einstein-Hilbert action of the gravitational field.

\subsection{The renormalized stress-energy tensor in $\mathrm {AdS}^4$}
\label{Sec.IIIc}

\textit{Mutatis mutandis}, the calculations of Sec.~\ref{Sec.IIIb} can be adapted to obtain the RSET of Stueckelberg electromagnetism in $\mathrm {AdS}^4$. We first determine the covariant Taylor series expansions \eqref{covTaylorSeries_WVector} and \eqref{covTaylorSeries_WScalar} from the expressions \eqref{Sol_G1VectorFn_AdS_a} and \eqref{Sol_G1ScalarFn_AdS_a} of the Hadamard Green functions. We have

\begin{widetext}

\begin{subequations}
\label{TaylorCoeff_in_AdS_WAetW}
\begin{eqnarray}
&&  s_{\mu \nu} = \Big\{ - (1/2) \, m^2 +(5/144) \, R \nonumber \\
&&  \hphantom{s_{\mu \nu} = }
   + \big[ (3/8) \, m^2 + (1/16) \, R \big] \big[ \ln(-R/24) + 2 \Psi(1/2+\lambda) + 2 \gamma \big] \nonumber \\
&&  \hphantom{s_{\mu \nu} = }
   + \big[ (1/8) \, m^2 - (1/48) \, R \big] \big[ \ln(-R/24) + 2 \Psi(1/2+\kappa) + 2 \gamma \big]
\Big\} \, g_{\mu \nu} , \label{TaylorCoeff_in_AdS_WA_a}\\
&&  s_{\mu \nu a b} = \Big\{ - (5/16) \, m^4 - (1/18) \, m^2 R + (119/51840) \, R^2 \nonumber \\
&&  \hphantom{s_{\mu \nu a b} = }
   + \big[ (5/48) \, m^4 + (11/288) \, m^2 R + (1/288) \, R^2 \big] \big[ \ln(-R/24) + 2 \Psi(1/2+\lambda) + 2 \gamma \big] \nonumber \\
&&  \hphantom{s_{\mu \nu a b} = }
   + \big[ (1/48) \, m^4 + (1/288) \, m^2 R - (1/864) \, R^2  \big] \big[ \ln(-R/24) + 2 \Psi(1/2+\kappa) + 2 \gamma \big]
\Big\} \, g_{\mu \nu} g_{a b} \nonumber \\
&&  \hphantom{s_{\mu \nu a b} } + \Big\{ (1/144) \, m^2 R + (1/5184) \, R^2 \nonumber \\
&&  \hphantom{s_{\mu \nu a b} = }
   - \big[ (1/24) \, m^4 - (1/288) \, m^2 R - (1/576) \, R^2 \big] \big[ \ln(-R/24) + 2 \Psi(1/2+\lambda) + 2 \gamma \big] \nonumber \\
&&  \hphantom{s_{\mu \nu a b} = }
   + \big[ (1/24) \, m^4 - (1/288) \, m^2 R - (1/1728) \, R^2  \big] \big[ \ln(-R/24) + 2 \Psi(1/2+\kappa) + 2 \gamma \big]
\Big\} \, g_{\mu (a|} g_{\nu |b)} \label{TaylorCoeff_in_AdS_WA_b}
\end{eqnarray}
and
\begin{eqnarray}
&&  w = - (1/2) \, m^2 + (7/72) \, R
   + \big[ (1/2) \, m^2 - (1/12) \, R \big] \big[ \ln(-R/24) + 2 \Psi(1/2+\kappa) + 2 \gamma \big],
\label{TaylorCoeff_in_AdS_W_c}\\
&&  w_{a b} = \Big\{ - (5/16) \, m^4 + (25/288) \, m^2 R - (29/5760) \, R^2 \nonumber \\
&&  \hphantom{s_{\mu \nu a b} = }
   + \big[ (1/8) \, m^4 - (1/48) \, m^2 R \big] \big[ \ln(-R/24) + 2 \Psi(1/2+\kappa) + 2 \gamma \big]
\Big\} g_{a b} . \label{TaylorCoeff_in_AdS W_d}
\end{eqnarray}
\end{subequations}
\end{widetext}
In order to simplify the previous expressions and, in particular, to eliminate the terms in $\tan (\pi \kappa)$ and in $\tan (\pi \lambda)$ occurring in the expressions \eqref{Sol_G1VectorFn_AdS_a} and \eqref{Sol_G1ScalarFn_AdS_a} of the Hadamard Green functions, we have used systematically the relation
\begin{align}
\label{DigammaFunction_et_tg}
& \Psi (n+1/2+z) + \Psi (n+1/2 - z) + \pi  \tan (\pi z) = \nonumber \\
&  \qquad 2 \Psi (1/2+ z) + 2 \, (1- \delta_{n 0}) \sum_{p=0}^{n-1} \frac{p+1/2}{(p+1/2)^2-z^2} \nonumber \\
&
\end{align}
(here $\delta_{n m}$ is the Kronecker delta) which is valid for $n \in \mathbb{N}$. This relation can be derived from the reflection formula \cite{AS65,Olver:2010:NHMF,NIST:DLMF}
\begin{align}
\label{DigammaFunction_ref}
\Psi (1-z)= \Psi (z) + \pi  \cot (\pi z)
\end{align}
\noindent making use of $\tan (\pi z)=-\cot [\pi (z+  1/2)]$ and of the recurrence formula \eqref{Psi_z_ppté1}. From the results \eqref{TaylorCoeff_in_AdS_WA_a}, \eqref{TaylorCoeff_in_AdS_WA_b} and \eqref{TaylorCoeff_in_AdS W_d}, we can write
\begin{widetext}
\begin{subequations}
\label{Tr_TaylorCoeff_in_AdS_WAetW}
\begin{eqnarray}
&&  s^{\phantom{\rho} \rho}_\rho = - 2 \, m^2 +(5/36) \, R \nonumber \\
&&  \hphantom{s^{\phantom{\rho} \rho}_\rho = }
   + \big[ (3/2) \, m^2 + (1/4) \, R \big] \big[ \ln(-R/24) + 2 \Psi(1/2+\lambda) + 2 \gamma \big] \nonumber \\
&&  \hphantom{s^{\phantom{\rho} \rho}_\rho = }
   + \big[ (1/2) \, m^2 - (1/12) \, R \big] \big[ \ln(-R/24) + 2 \Psi(1/2+\kappa) + 2 \gamma \big] , \label{Tr_TaylorCoeff_in_AdS_WA_a}\\
&&  s^{\phantom{\rho \tau} \rho \tau}_{\rho \tau} =  - (5/4) \, m^4 - (11/72) \, m^2 R + (1/90) \, R^2 \nonumber \\
&&  \hphantom{s_{\mu \nu a b} = }
   + \big[ (3/16) \, m^2 R + (1/32) \, R^2  \big] \big[ \ln(-R/24) + 2 \Psi(1/2+\lambda) + 2 \gamma \big] \nonumber \\
&&  \hphantom{s_{\mu \nu a b} = }
   + \big[ (1/2) \, m^4 - (1/48) \, m^2 R - (1/96) \, R^2 \big] \big[ \ln(-R/24) + 2 \Psi(1/2+\kappa) + 2 \gamma \big] \label{Tr_TaylorCoeff_in_AdS_WA_b}
\end{eqnarray}
and
\begin{eqnarray}
&&  w^{\phantom{\rho} \rho}_\rho = - (5/4) \, m^4 + (25/72) \, m^2 R - (29/1440) \, R^2 \nonumber \\
&&  \hphantom{w^{\phantom{\rho} \rho}_\rho = }
   + \big[ (1/2) \, m^4 - (1/12) \, m^2 R \big] \big[ \ln(-R/24) + 2 \Psi(1/2+\kappa) + 2 \gamma \big] . \label{Tr_TaylorCoeff_in_AdS_W_c}
\end{eqnarray}
\end{subequations}
\end{widetext}
It should be noted that the constraints \eqref{Relation_coefTaylorSeries_W_1} and \eqref{Relation_coefTaylorSeries_WA_W} are satisfied by these coefficients.

By inserting now the expressions \eqref{Tr_TaylorCoeff_in_AdS_WA_a} and \eqref{Tr_TaylorCoeff_in_AdS_WA_b} into \eqref{SET_MaximallySym_TR1} taking into account the geometrical term \eqref{TRv1A_MaximallySym} as well as the geometrical ambiguities \eqref{AmbGEN_MaximallySym} and \eqref{AmbGEN_MaximallySym_TR1}, we have for the trace of the RSET
\begin{widetext}
\begin{eqnarray}
\label{SET_AdS4_TR1}
&& 8\pi^2 \langle 0 | \widehat{T}^{\phantom{\rho} \rho}_{\rho} | 0 \rangle_\mathrm{ren \,\,\mathrm {AdS}^4} = (\alpha + 9/4) \, m^4 + (\beta + 5/24) m^2 R - (11/1440) \, R^2
\nonumber \\
&& \phantom{8\pi^2 \langle 0 | \widehat{T}^{\phantom{\rho} \rho}_{\rho} | 0 \rangle_\mathrm{ren \,\,\mathrm {AdS}^4} = } - \big[ (3/2) \, m^4 + (1/4) \, m^2 R \big] \big[ \ln(-R/(24M^2)) + 2 \Psi(1/2+\lambda) + 2 \gamma \big].
\end{eqnarray}
Finally, by requiring the vanishing of this expression in the flat-space limit, i.e. for $R \to 0$, we can fix the renormalization mass $M$ (after absorption of the term $2 \gamma$) and the coefficient $\alpha$. We then obtain $M=m/\sqrt{2}$ and $\alpha = -9/4$ which leads to

\begin{eqnarray}
\label{SET_AdS4_TR1_def}
&& 8\pi^2 \langle 0 | \widehat{T}^{\phantom{\rho} \rho}_{\rho} | 0 \rangle_\mathrm{ren \,\,\mathrm {AdS}^4} = ( \beta +5/24 ) \, m^2 R - (11/1440) \, R^2
\nonumber \\
&& \phantom{8\pi^2 \langle 0 | \widehat{T}^{\phantom{\rho} \rho}_{\rho} | 0 \rangle_\mathrm{ren \,\,\mathrm {AdS}^4} = } - \big[ (3/2) \, m^4 + (1/4) \, m^2 R \big] \big[ \ln(-R/(12m^2)) + 2 \Psi(1/2+\lambda) \big] .
\end{eqnarray}

\end{widetext}

\subsection{Remarks concerning the zero-mass limit of the renormalized stress-energy tensor}
\label{Sec.IIId}

Because Stueckelberg massive electromagnetism is a $U(1)$ gauge theory which generalizes Maxwell's theory, we could naively expect to recover, by considering the zero-mass limit of the previous results, the usual trace anomaly for Maxwell's theory given by [see, e.g., Eq.~(130) in Ref.~\cite{Belokogne:2015etf} or Eq.~(3.25) in Ref.~\cite{Brown:1986tj}]
\begin{eqnarray}
\label{SET_TrAnom}
&& 8\pi^2 \langle 0 | \widehat{T}^A{}^{\phantom{\rho} \rho}_{\rho} | 0 \rangle_\mathrm{ren} = 2 v^A_1{}^{\phantom{\rho} \rho}_{\rho} - 4 v_1
\end{eqnarray}
which reduces, in a maximally symmetric gravitational background, to
\begin{eqnarray}
\label{SET_MS_TRAmu_m=0_MS}
&& 8\pi^2 \langle 0 | \widehat{T}^A{}^{\phantom{\rho} \rho}_{\rho} | 0 \rangle_\mathrm{ren} = - (31/2160) \, R^2.
\end{eqnarray}
In fact, this is not the case. In $\mathrm {dS}^4$, for $m^2 \to 0$, Eq.~\eqref{SET_dS4_TR1_def} provides
\begin{eqnarray}
\label{SET_dS4_TR_def_m=0}
&& 8\pi^2 \langle 0 | \widehat{T}^{\phantom{\rho} \rho}_{\rho} | 0 \rangle_\mathrm{ren \,\,\mathrm {dS}^4} =   (19/1440) \, R^2,
\end{eqnarray}
while, in $\mathrm {AdS}^4$, for $m^2 \to 0$, we obtain from Eq.~\eqref{SET_AdS4_TR1_def}
\begin{eqnarray}
\label{SET_AdS4_TR_def_m=0}
&& 8\pi^2 \langle 0 | \widehat{T}^{\phantom{\rho} \rho}_{\rho} | 0 \rangle_\mathrm{ren \,\,\mathrm {AdS}^4} =   -(11/1440) \, R^2.
\end{eqnarray}
This ``discontinuity'' is not really surprising. Indeed, as we have already noted in Ref.~\cite{Belokogne:2015etf}, in an arbitrary spacetime, due to the contribution of the auxiliary scalar field $\Phi$, the full RSET of Stueckelberg electromagnetism never permits us to recover the RSET of Maxwell's theory. This can be alternatively interpreted by noting that the presence of a mass term in the Stueckelberg theory breaks the conformal invariance of Maxwell's theory. To circumvent this difficulty, we have proposed in Ref.~\cite{Belokogne:2015etf} to split the RSET of Stueckelberg electromagnetism into two separately conserved RSETs, a contribution directly associated with the vector field $A_\mu$ and another one corresponding to the scalar field $\Phi$ (see also our discussion in Sec.~\ref{Sec.IIIa}), the
zero-mass limit of the first contribution reducing to the RSET of Maxwell's electromagnetism. Even if we consider that this separation is rather artificial because, in our opinion, only the full RSET is physically relevant, the auxiliary scalar field $\Phi$ playing the role of a kind of ghost field, it is interesting to test our proposal which has nevertheless provided correct results in the context of the Casimir effect (see Sec.~V of Ref.~\cite{Belokogne:2015etf}).

In $\mathrm {dS}^4$, we can obtain the separated RSETs associated with the massive vector field $A_\mu$ and the massive scalar field $\Phi$ which vanish in the flat-space limit by inserting into Eqs.~\eqref{SET_MaximallySym_TR2_A} and \eqref{SET_MaximallySym_TR2_Phi} the Taylor coefficients \eqref{Tr_TaylorCoeff_in_dS_WA_a}, \eqref{Tr_TaylorCoeff_in_dS_WA_b} and \eqref{TaylorCoeff_in_dS_W_c} and by moreover taking into account the geometrical ambiguities \eqref{AmbGEN_MaximallySym_TR2_A}, \eqref{AmbGEN_MaximallySym_TR2_Phi} and \eqref{AmbGEN_MaximallySym}. We have
\begin{widetext}
\begin{eqnarray}
\label{SET_dS4_TRAmu_def}
&& 8\pi^2 \langle 0 | \widehat{T}^A{}^{\phantom{\rho} \rho}_{\rho} | 0 \rangle_\mathrm{ren \,\,\mathrm {dS}^4} = ( \beta_A + 13/18 ) \, m^2 R + (59/2160) \, R^2
\nonumber \\
&& \phantom{8\pi^2 \langle 0 | \widehat{T}^A{}^{\phantom{\rho} \rho}_{\rho} | 0 \rangle_\mathrm{ren \,\,\mathrm {dS}^4} = } - \big[ (3/2) \, m^4 + (1/4) \, m^2 R \big] \big[ \ln(R/(12m^2)) + \Psi(5/2+\lambda) + \Psi(5/2-\lambda) \big]
\nonumber \\
&& \phantom{8\pi^2 \langle 0 | \widehat{T}^A{}^{\phantom{\rho} \rho}_{\rho} | 0 \rangle_\mathrm{ren \,\,\mathrm {dS}^4} = } + \big[ (1/2) \, m^4 - (1/12) \, m^2 R \big] \big[ \ln(R/(12m^2)) +  \Psi(5/2+\kappa) +  \Psi(5/2 - \kappa) \big]
\end{eqnarray}
for the RSET associated with the massive vector field $A_\mu$ and
\begin{eqnarray}
\label{SET_dS4_TRPhi_def}
&& 8\pi^2 \langle 0 | \widehat{T}^\Phi{}^{\phantom{\rho} \rho}_{\rho} | 0 \rangle_\mathrm{ren \,\,\mathrm {dS}^4} = (\beta_\Phi - 5/36) \, m^2 R + (29/4320) \, R^2
\nonumber \\
&& \phantom{8\pi^2 \langle 0 | \widehat{T}^\Phi{}^{\phantom{\rho} \rho}_{\rho} | 0 \rangle_\mathrm{ren \,\,\mathrm {dS}^4} = } - \big[ (1/2) \, m^4 - (1/12) \, m^2 R \big] \big[ \ln(R/(12m^2)) +  \Psi(3/2+\kappa) +  \Psi(3/2 - \kappa) \big]
\end{eqnarray}
\end{widetext}
for the RSET associated with the massive scalar field $\Phi$. It is worth pointing out that the sum of these two RSETs coincides with the RSET given by Eq.~\eqref{SET_dS4_TR1_def}. This is a trivial consequence of Eq.~\eqref{Psi_z_ppté1}. We can moreover note that the RSET \eqref{SET_dS4_TRPhi_def} associated with the scalar field $\Phi$ is nothing else than the result derived by Bunch and Davies in Ref.~\cite{Bunch:1978yq} (see also Refs.~\cite{Dowker:1975tf,Bernard:1986vc,Tadaki:1988cn}. If we now consider the limit $m^2 \to 0$ of the RSET \eqref{SET_dS4_TRAmu_def} associated with the vector field $A_\mu$, we obtain
\begin{eqnarray}
\label{SET_dS4_TRAmu_m=0}
&& 8\pi^2 \langle 0 | \widehat{T}^A{}^{\phantom{\rho} \rho}_{\rho} | 0 \rangle_\mathrm{ren \,\,\mathrm {dS}^4} =  (59/2160) \, R^2.
\end{eqnarray}
Here, we do not recover the usual trace anomaly \eqref{SET_MS_TRAmu_m=0_MS} of Maxwell's electromagnetism. In fact, this can be explained easily. Indeed, in Secs.~IV~C~4 and IV~D. of Ref.~\cite{Belokogne:2015etf}, we have constructed the RSET associated with the massive vector field $A_\mu$ which reduces, in the zero-mass limit, to the RSET of Maxwell's electromagnetism by assuming tacitely the regularity for $m^2 \to 0$ of the Taylor coefficients involved in the calculations and, in particular, of the coefficient $w$. This assumption is certainly valid for a large class of spacetimes but, in the context of field theory in $\mathrm {dS}^4$, it is not founded. For example, from Eq.~\eqref{TaylorCoeff_in_dS_W_c}, we can show that $w \sim R^2/(48 \, m^2)$ for $m^2 \to 0$. We here encounter the well-known infrared divergence problem which plagues quantum field theories in de Sitter spacetime (see, e.g., Ref~\cite{Allen:1985ux} and references therein).

In $\mathrm {AdS}^4$, we can obtain the separated RSETs associated with the massive vector field $A_\mu$ and the massive scalar field $\Phi$ which vanish in the flat-space limit by inserting into Eqs.~\eqref{SET_MaximallySym_TR2_A} and \eqref{SET_MaximallySym_TR2_Phi} the Taylor coefficients \eqref{Tr_TaylorCoeff_in_AdS_WA_a}, \eqref{Tr_TaylorCoeff_in_AdS_WA_b} and \eqref{TaylorCoeff_in_AdS_W_c} and by moreover taking into account the geometrical ambiguities \eqref{AmbGEN_MaximallySym_TR2_A}, \eqref{AmbGEN_MaximallySym_TR2_Phi} and \eqref{AmbGEN_MaximallySym}. We have
\begin{widetext}
\begin{eqnarray}
\label{SET_AdS4_TRAmu_def}
&& 8\pi^2 \langle 0 | \widehat{T}^A{}^{\phantom{\rho} \rho}_{\rho} | 0 \rangle_\mathrm{ren \,\,\mathrm {AdS}^4} = ( \beta_A +7/18 ) \, m^2 R - (31/2160) \, R^2
\nonumber \\
&& \phantom{8\pi^2 \langle 0 | \widehat{T}^A{}^{\phantom{\rho} \rho}_{\rho} | 0 \rangle_\mathrm{ren \,\,\mathrm {AdS}^4} = } - \big[ (3/2) \, m^4 + (1/4) \, m^2 R \big] \big[ \ln(-R/(12m^2)) + 2 \Psi(1/2+\lambda) \big]
\nonumber \\
&& \phantom{8\pi^2 \langle 0 | \widehat{T}^A{}^{\phantom{\rho} \rho}_{\rho} | 0 \rangle_\mathrm{ren \,\,\mathrm {AdS}^4} = } + \big[ (1/2) \, m^4 - (1/12) \, m^2 R \big] \big[ \ln(-R/(12m^2)) + 2 \Psi(1/2+\kappa) \big]
\end{eqnarray}
for the RSET associated with the massive vector field $A_\mu$ and
\begin{eqnarray}
\label{SET_AdS4_TRPhi_def}
&& 8\pi^2 \langle 0 | \widehat{T}^\Phi{}^{\phantom{\rho} \rho}_{\rho} | 0 \rangle_\mathrm{ren \,\,\mathrm {AdS}^4} = ( \beta_\Phi - 13/72 ) \, m^2 R + (29/4320) \, R^2
\nonumber \\
&& \phantom{8\pi^2 \langle 0 | \widehat{T}^\Phi{}^{\phantom{\rho} \rho}_{\rho} | 0 \rangle_\mathrm{ren \,\,\mathrm {AdS}^4} = } - \big[ (1/2) \, m^4 - (1/12) \, m^2 R \big] \big[ \ln(-R/(12m^2)) + 2 \Psi(1/2+\kappa) \big]
\end{eqnarray}
\end{widetext}
for the RSET associated with the massive scalar field $\Phi$. We can note that the sum of these two RSETs coincides with the RSET given by Eq.~\eqref{SET_AdS4_TR1_def} and that the RSET \eqref{SET_AdS4_TRPhi_def} associated with the scalar field $\Phi$ is in agreement with the result derived by Camporesi and Higuchi in Ref.~\cite{Camporesi:1992wn} (see also Ref.~\cite{Kent:2014nya}). If we now consider the limit $m^2 \to 0$ of the RSET \eqref{SET_AdS4_TRAmu_def} associated with the vector field $A_\mu$, we obtain
\begin{eqnarray}
\label{SET_AdS4_TRAmu_m=0}
&& 8\pi^2 \langle 0 | \widehat{T}^A{}^{\phantom{\rho} \rho}_{\rho} | 0 \rangle_\mathrm{ren \,\,\mathrm {AdS}^4} =  - (31/2160) \, R^2.
\end{eqnarray}
In $\mathrm {AdS}^4$, we recover the usual trace anomaly \eqref{SET_MS_TRAmu_m=0_MS} of Maxwell's electromagnetism.

\section{Conclusion}
\label{Sec.IV}

In the present article, by focusing on Hadamard vacuum states, we have first constructed the various two-point functions associated with Stueckelberg massive electromagnetism in de Sitter and anti-de Sitter spacetimes. Then, from the general formalism developed in Ref.~\cite{Belokogne:2015etf}, we have obtained an exact analytical expression for the vacuum expectation
value of the RSET of the massive vector field propagating in de Sitter and anti-de Sitter spacetimes. It is worth pointing out that these results have been obtained by working in the unique gauge for which the machinery of Hadamard renormalization can be used (i.e., $\xi=1$). However, in the literature, Stueckelberg massive electromagnetism is often considered for an arbitrary gauge parameter $\xi$, i.e., by describing the massive vector field from the action \eqref{Action_Stueck_Amu_xi} instead of the action \eqref{Action_Stueck_Amu_xi=1}. Here, we refer to the articles by Fr{\"o}b and Higuchi \cite{Frob:2013qsa} who consider the massive vector field in de Sitter spacetime and by Janssen and Dullemond \cite{Janssen:1986fz} who work in anti-de Sitter spacetime. The propagators constructed in this context are $\xi$ dependent but have good physical properties and, even if they do not have a Hadamard-type singularity at short distance (for $\xi \not= 1$), they reproduce  in the flat-space limit the standard Minkowski two-point functions given by Itzykson and Zuber in Ref.~\cite{ItzyksonZuber}. It would be interesting to analyze the physical content of these propagators by constructing their associated RSETs. Due to the expected gauge independence of the RSET, it is quite likely that the results obtained will be identical to those given in the previous section but a proof of this claim certainly requires a careful study. We plan to provide it in the near future.

\begin{acknowledgments}

We wish to thank Yves D\'ecanini and Mohamed Ould El Hadj for various discussions, Philippe Spindel for providing us with a copy of Ref.~\cite{SchomblondSpindel76} and the ``Collectivit\'e Territoriale de Corse'' for its support through the COMPA project.

\end{acknowledgments}

%
%
%
%
%
%
%
%

\bibliography{Stueckelberg_dS_AdS}

\end{document}